\documentclass[journal]{IEEEtran}

\usepackage{multirow}
\usepackage{subfigure}

\usepackage{graphicx}
\usepackage{amsmath}
\usepackage{amsfonts}
\usepackage{algorithm}
\usepackage{algorithmic}
\usepackage{makecell}
\usepackage{booktabs}
\setcellgapes{3pt}

\setcellgapes{3pt}

\usepackage[justification=centering]{caption}
\usepackage{color}
\usepackage{easyReview}

\usepackage[english]{babel}

\usepackage{cite}
\usepackage{verbatim}

\ifCLASSINFOpdf

\else

\fi

\begin{document}

\title{SLM, LLM or Agentic AI? Toward Intelligent UAV-Enabled WPT Systems in Low-Altitude Economy Networks}

\author{Feibo Jiang, \textit{Senior Member, IEEE}, Li Dong, Lei Mao, Kezhi Wang, \textit{Senior Member, IEEE}, Xianbin Wang, \textit{Fellow, IEEE}, Abbas Jamalipour, \textit{Fellow, IEEE}.% <-this % stops a space
\thanks{This work was supported in part by the National Natural Science Foundation
	of China under Grants 62572184, and 41604117; in part by the
	Natural Science Foundation of Hunan Province under Grants 2024JJ5270 and
	2025JJ50365; in part by the Changsha Natural Science Foundation under
	Grants kq2402098 and kq2402162 (Corresponding author: Li Dong.)}
\thanks{Feibo Jiang (jiangfb@hunnu.edu.cn) is with the Hunan Provincial Key
	Laboratory of Intelligent Computing and Language Information Processing
	and the Yuelushan Digital Intelligence Laboratory (Artificial Intelligence and
	International Communication), Hunan Normal University, Changsha 410081,
	China.}% <-this % stops a space
\thanks{Li Dong (Dlj2017@hunnu.edu.cn) is with the School of Computer Science,
	Hunan University of Technology and Business, Changsha 410205, China, and
	also with the Xiangjiang Laboratory, Changsha 410205, China.}
\thanks{Lei Mao (202520294367@hunnu.edu.cn) is with School of Information
Science and Engineering, Hunan Normal University, Changsha 410081, China.}

\thanks{Kezhi Wang (Kezhi.Wang@brunel.ac.uk) is with the Department of Computer
	Science, Brunel University London, London UB8 3PH, UK.}
\thanks{Xianbin Wang (xianbin.wang@uwo.ca)  is with the Department of Electrical and Computer Engineering,
Western University, London N6A 5B9, Canada.}  

\thanks{Abbas Jamalipour (a.jamalipour@ieee.org) is with the School of Electrical and Computer Engineering, University of Sydney, Australia, and with the Graduate School of Information Sciences, Tohoku University, Japan.}
}

\markboth{Submitted for Review}%
{Shell \MakeLowercase{\textit{et al.}}: Bare Demo of IEEEtran.cls for IEEE Journals}

\maketitle

\begin{abstract}
Unmanned Aerial Vehicles (UAVs) have become key enabling platforms for low-altitude economic networks, yet achieving efficient and adaptive optimization under resource-constrained and dynamic environments remains challenging. This paper investigates language models for UAV-enabled Wireless Power Transfer (WPT) systems. First, a lightweight Small Language Model (SLM)-based solution is developed using a pre-trained BERT backbone, enhanced UAV embeddings and contextual features, a geometry-aware path decoder, and ensemble inference to achieve low complexity, low latency, and high energy efficiency. Second, an Agentic AI-based framework is designed to exploit the reasoning and interactive capabilities of Large Language Models (LLMs). It integrates four collaborative agents—Initializer, Actor, Critic, and Reflector—to form a closed loop of generation, optimization, evaluation, and reflection for iterative UAV path and energy optimization. Finally, simulations compare the SLM-, LLM-, and Agentic AI-based approaches.

\end{abstract}


\IEEEpeerreviewmaketitle

\section{Introduction}
\label{sec:introduction}
\subsection{\textcolor{black}{Background}}
In recent years, the Low-Altitude Economy Network (LAENet) has emerged as a new frontier of future information infrastructure and economic growth. It is gradually evolving into a multi-layered network that integrates transportation, energy supply, air–ground collaboration, and intelligent services. Within LAENet, Unmanned Aerial Vehicle (UAV) has become a core node supporting network operation and value realization due to its high mobility, rapid deployment, and wide-area coverage capabilities. However, UAV systems also face multi-dimensional challenges related to limited computing, communication, and energy resources, which place higher demands on their ability to operate over extended periods.

Energy consumption is a critical bottleneck that constrains both the flight endurance and service capabilities of UAVs. Due to payload and size limitations, the onboard battery capacity of UAVs is inherently restricted, making it difficult to sustain long-duration missions and large-scale networked applications \cite{11173659}. To address this issue, Wireless Power Transfer (WPT) technology has been integrated into UAV systems \cite{10879146}. WPT enables UAVs to receive energy replenishment during flight and, at the same time, function as energy nodes that provide wireless power to ground terminals or other devices. This approach effectively overcomes the limitations of traditional battery-powered modes. By extending individual UAV flight time and enabling sustained operation of multi-UAV and terminal networks, WPT significantly improves overall task completion rates and resource utilization efficiency.

\subsection{\textcolor{black}{Motivation}}
Path planning plays a crucial role in UAV-enabled WPT system architectures. During mission execution, UAVs must simultaneously consider multiple factors, including flight energy consumption, energy replenishment locations, communication coverage, and mission delay. These interrelated requirements make path planning a highly complex optimization problem. On one hand, dynamic environmental changes, uncertain energy transfer efficiency, and diverse task demands introduce significant uncertainty and high-dimensional coupling into the UAV path optimization process. On the other hand, traditional methods, which mainly rely on heuristic algorithms \cite{9523743} and convex optimization algorithms \cite{9276730}, can improve path selection and energy efficiency to some extent but often suffer from slow convergence, high computational complexity, and limited adaptability to dynamic environments. Achieving energy-efficient, intelligent, and robust path planning remains a key research focus and a major challenge in this field.

In recent years, the rapid progress of Large Language Models (LLMs) has introduced promising opportunities for enhancing the optimization of UAV-enabled WPT systems, owing to their powerful capabilities in complex reasoning, strategic planning, and cross-modal understanding \cite{11299527}. By enabling natural language interaction, knowledge transfer, and adaptive optimization, LLMs can intelligently interpret dynamic task environments and support efficient decision-making. This makes them a promising tool for enhancing multi-UAV collaboration and energy-efficient path planning \cite{11370829}. However, the significant training and inference costs of LLMs pose critical challenges. Effectively integrating LLM-based intelligent reasoning capabilities while maintaining system energy efficiency and accommodating resource constraints remains a key issue for future research.

\subsection{\textcolor{black}{Contributions}}

Against this backdrop, this paper proposes and systematically investigates three intelligent solution frameworks for UAV-enabled WPT systems in LAENets, aiming to strike a balance between model performance and resource consumption. The contributions of this study are summarized as follows.

\subsubsection{\textcolor{black}{SLM-Based Lightweight Optimization}}
\textcolor{black}{To address the resource constraints of energy-efficient UAV-enabled WPT systems, we propose a lightweight SLM-based solution. Built on a pre-trained BERT backbone, it integrates enhanced UAV embeddings, contextual features, a geometry-aware path decoder, ensemble inference, and reinforcement learning-based optimization. These designs improve path-geometry modeling and optimization performance. The resulting SLM-based approach achieves low computational complexity, high energy efficiency, and low inference latency, making it suitable for real-time and resource-constrained UAV operations in LAENets.}

\subsubsection{\textcolor{black}{LLM- and Agentic AI-Based Reasoning}}
\textcolor{black}{To handle the uncertainty and dynamic coupling between path planning and energy consumption, we further explore the reasoning and interactive capabilities of LLMs through an Agentic AI framework. This framework includes four cooperative agents, namely Initializer, Actor, Critic, and Reflector, forming a closed loop of solution generation, optimization, evaluation, and feedback refinement. In addition, the Critic uses the Model Context Protocol (MCP) to invoke external tools for constraint verification and local path search, thereby improving generalization and adaptability in dynamic environments.}

\subsubsection{\textcolor{black}{Cross-Paradigm Evaluation and Complementarity}}
\textcolor{black}{Through theoretical analysis and simulations, we compare the SLM-, LLM-, and Agentic AI-based schemes in terms of energy efficiency, path quality, and computational complexity. Results show that SLM has clear advantages in resource-constrained scenarios, whereas Agentic AI provides stronger decision-making intelligence and robustness in dynamic environments. Further analysis reveals their complementary roles and applicability boundaries in LAENets, offering insights into the integration of lightweight modeling and intelligent reasoning for future wireless systems.}

\subsection{\textcolor{black}{Organization}}
The remainder of this paper is organized as follows.
Section~\ref{sec:related} reviews the related work.
Section~\ref{sec:system_model} presents the system model and problem formulation.
Section~\ref{sec:slm} introduces the SLM-based solution framework, while Section~\ref{sec:llm} elaborates on the LLM and Agentic AI–based solution framework.
Section~\ref{sec:experiments} reports and analyzes the experimental results.
Finally, Section~\ref{sec:conclusion} concludes the paper.

\section{Related Work}
\label{sec:related}

\subsection{Energy-Efficient UAV-Enabled WPT Systems}
Yang et al. \cite{yang2024deep} proposed a Dueling Double Deep Q-Network (Dueling DDQN)-based UAV-IoT system that jointly optimizes trajectory, power allocation, and scheduling in a Simultaneous Wireless Information and Power Transfer (SWIPT) scenario. Dong et al. \cite{dong2025attention} proposed an Attention-based UAV Trajectory Optimization (AUTO) framework for WPT-assisted IoT systems, achieving multi-UAV trajectory and deployment optimization. By integrating the Attention-based Trajectory Optimization Model (ATOM), the framework enhances both energy efficiency and stability, effectively addressing trajectory optimization challenges in large-scale IoT networks.
Muy et al. \cite{muy2024trajectory} proposed a grid-world Deep Reinforcement Learning (DRL)-based UAV-WPT system that jointly optimizes trajectory, beamforming, and power allocation. Kim et al. \cite{kim2024joint} proposed a multi-UAV WPT communication network that jointly optimizes trajectory, scheduling, and power allocation. The method enhances minimum uplink throughput under energy neutrality and mobility constraints while mitigating interference in multi-UAV scenarios.

\subsection{UAV System Optimization Based on DRL}
Ning et al. \cite{10239498} proposed a DRL-based UAV trajectory optimization algorithm, in which each service provider’s UAV makes flight decisions solely based on local observations. The objective is to simultaneously minimize the short-term computational cost of ground users and the long-term computational cost of UAVs under conditions of incomplete information.
Su et al. \cite{su2024deep} proposed a dual-UAV secure communication system based on the Deep Q-Networks (DQN). This system achieves joint optimization of trajectory, power, and user scheduling, thereby improving the secrecy rate and convergence speed while effectively addressing physical-layer security and anti-eavesdropping challenges in UAV communications.
Li et al. \cite{li2025deep} proposed a Soft Actor-Critic for Covert Communication and Charging (SAC-CC) system, which performs joint optimization of UAV trajectory and bandwidth allocation.

\begin{comment}
	内容...

\subsection{UAV System Optimization Based on Transformer}
Chen et al. \cite{chen2024transformer} proposed a Transformer-based deep Multi-Agent Reinforcement Learning (T-MARL) approach, which enables joint trajectory planning and area coverage optimization. This method improves coverage and fairness while effectively addressing the scalability challenges of UAVs in large-scale cooperative coverage scenarios.
Wang et al. \cite{wang2024enhanced} proposed an Enhanced Transformer-based UAV-to-UAV Tracking (ETUT) framework, which achieves more robust inter-UAV target tracking through interference suppression and trajectory prediction. This framework improves tracking accuracy and robustness, effectively addressing the problem of target loss in complex environments.
Souli et al. \cite{souli2025onboard} proposed a multitask system based on Transformer and Reservoir Computing (RC), which enables UAV state estimation and trajectory prediction with real-time operation on onboard platforms. This approach improves prediction accuracy and robustness, effectively addressing the challenges of reliable UAV monitoring and prediction in dynamic environments.
\end{comment}
\subsection{UAV System Optimization Based on LLM}
Samma et al. \cite{samma2025uav} proposed an LLM-based UAV visual path planning system that enables autonomous navigation in indoor environments where GPS signals are unavailable through visual perception. Jiang et al. \cite{jiang2025agentic} proposed an Agentic AI-based framework for multi-UAV trajectory optimization in LAENet. Integrating Agentic RAG, Transformer and Mamba, and Trajectory-Group Relative Policy Optimization (T-GRPO), the framework enhances energy efficiency, convergence, and generalization in dynamic environments.
Yin et al. \cite{yin2025trajectory} proposed a Multimodal Large Language Model (M2LLM)-driven DRL framework for trajectory design and beamforming optimization in UAV networks. This approach improves spectrum efficiency, energy efficiency, and convergence speed, effectively addressing the joint optimization challenges of UAVs in complex environments.

\textcolor{black}{Therefore, although existing studies have achieved significant progress in energy-efficient optimization and task scheduling, current approaches still exhibit notable limitations in complex and dynamic scenarios. Specifically, traditional optimization struggles with high-dimensional, multi-objective search spaces; DRL still faces challenges in generalization and stability; and existing LLM-based studies have yet to establish deployable multi-agent reasoning frameworks. These limitations reveal critical gaps in the current literature and further underscore the necessity of developing a unified SLM–LLM–Agentic AI framework. Consequently, a systematic comparison between SLM-based methods and LLM-driven Agentic AI paradigms is of great significance for clarifying their complementary strengths, guiding model selection and optimization under different scenarios, and advancing the intelligent evolution of UAV-enabled WPT systems.}

\begin{table}[h]
	\centering
	\caption{\textcolor{black}{Comparison with previous works}}
	\label{tab:related_work_comparison}
	\renewcommand\arraystretch{1.25}
    {\color{black}
	\begin{tabular}{|
>{\centering\arraybackslash}p{1.5cm}|
>{\centering\arraybackslash}p{1.2cm}|
>{\centering\arraybackslash}p{1.2cm}|
>{\centering\arraybackslash}p{1.2cm}|
>{\centering\arraybackslash}p{1.5cm}|} 
		\hline
		\textbf{Work} & \textbf{UAV} & \textbf{WPT} & \textbf{DRL/RL} & \textbf{LLM / Agentic AI} \\ 
		\hline
		
		\cite{yang2024deep, dong2025attention, muy2024trajectory, kim2024joint} 
		& $\checkmark$ & $\checkmark$ & $\checkmark$ &  \\ 
		\hline
		
		\cite{10239498, su2024deep, li2025deep} 
		& $\checkmark$ &  & $\checkmark$ &  \\ 
		\hline
		
		\cite{samma2025uav, jiang2025agentic, yin2025trajectory} 
		& $\checkmark$ &  & $\checkmark$ & $\checkmark$ \\ 
		\hline
		
		\textbf{Our Work} 
		& $\checkmark$ & $\checkmark$ & $\checkmark$ & $\checkmark$ \\ 
		
		\hline
	\end{tabular}
    }
\end{table}

\section{System Model and Problem Formulation }
\label{sec:system_model}

\begin{figure}[htbp]
\centering
\includegraphics[width=9cm]{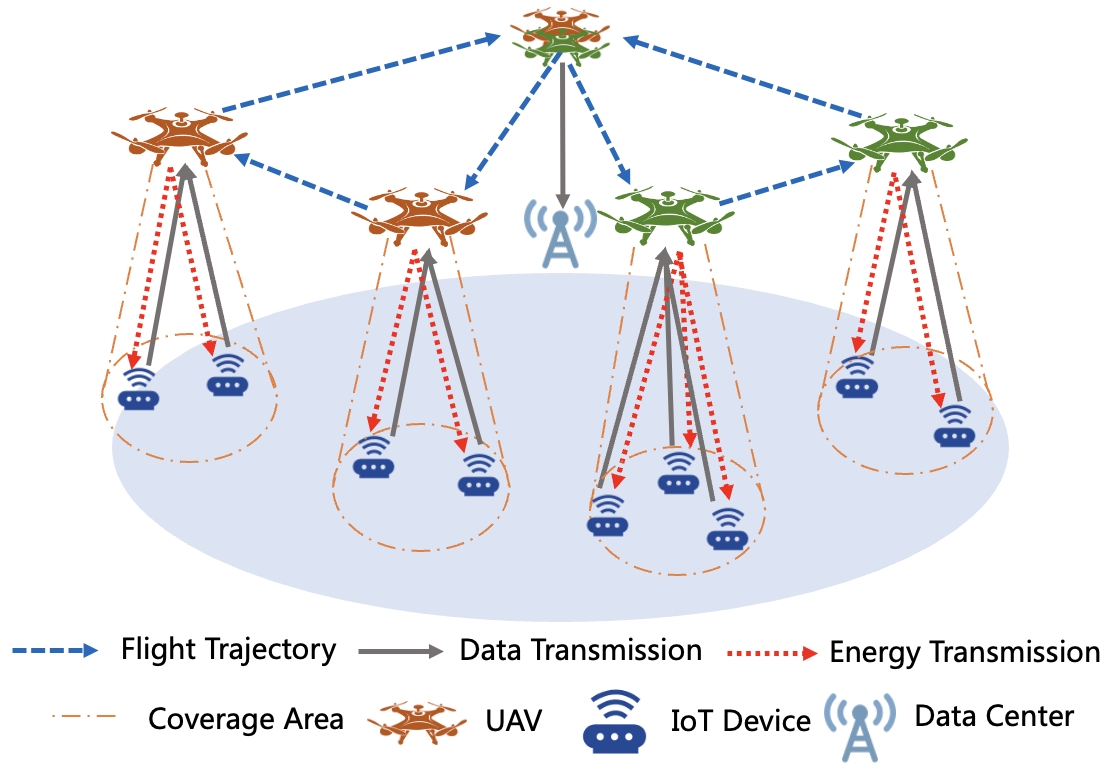}
\caption{\textcolor{black}{UAV-enabled WPT System with Two UAVs and Their Flight Paths for IoT Data Collection.}}
\label{fig:scenario}
\end{figure}

As illustrated in Fig. \ref{fig:scenario}, the UAV-enabled WPT system consists of $N$  IoT Devices (IoTDs), a data center, and $m$ UAVs with half-duplex access points, which can transmit power to the IoTDs and collect data from IoTDs by Time Division Duplexing (TDD) mode.
The $N$ IoTDs are denoted as a set of $\mathcal{N}=\left\{ 1,2,...,N \right\}$. 
We assume that the GPS is equipped in each IoTD and the position ($x_i, y_i$) of the $i$-th IoTD is fixed and known. The $i$-th IoTD has $D_{i}$ data to be collected. The set of $m$ UAVs is denoted as $\mathcal{M}=\left\{ 1,2,...,m \right\}$, and each UAV has limited data storage capacity $C_{\max}$ and energy capacity $E_{\max}$. The flight height of the UAV is set to $H_z$ meters. The data of each IoTD could be collected by one UAV at one time, so the association $a_{ij}$ at the $t$-th time step can be expressed by:
\begin{equation}
\label{eq:c_uav_collect}
a_{ij}[t]=\left\{0,1\right\}, \forall i\in\mathcal{N}, \forall j\in\mathcal{M}
\end{equation}
where $a_{ij}[t]=1$ means the $j$-th UAV is collecting the data from the $i$-th IoTD at the $t$-th time step, and $a_{ij}[t]=0$ otherwise. 
Assuming each IoTD can only be serviced once, then:
\begin{equation}
\label{eq:c_iot_transfer}
\sum_{j=1}^{m}\sum_{t=1}^{s_j} a_{ij}[t]=1.
\end{equation}
where ${s_j}$ is the total number of time steps.

\subsection{Communication Model}
Considering obstacles in the system, the simplified Line-of-Sight (LoS)-dominant channel model can be inaccurate in practice  \cite{2020prob-communication}.
In order to characterize the UAV-to-IoTD channel state more accurately, we adopt the probabilistic LoS channel model \cite{2014prob_los}, where LoS probability is a function of the elevation angle between UAV and IoTDs. A generalized logistic function is introduced to approximate the LoS probability as follows \cite{2020prob-communication}:
\begin{equation}
\label{eq:prob_los}
p_{ij}^{L}[t]=B_3+\frac{B_4}{1+e^{-(B_1+B_2 \theta_{ij}[t])}}
\end{equation}
where $\theta_{ij}[t]$ denotes the elevation angle between the $j$-th UAV and the $i$-th IoTD in time step $t$. $B_1,B_2,B_3,B_4$ are environment-related constants. Then, the corresponding non-LoS (NLoS) probability is obtained as $p^{N}_{ij}[t]=1-p_{ij}^{L}[t]$. Next, the large-scale channel power gain between the $i$-th IoTD and the $j$-th UAV in time step $t$, including both the path loss and NLoS attenuation, can be approximated as
\begin{equation}
\label{eq:prob_gain}
h_{ij}[t]=\left\{
\begin{array}{ll}
\beta_0(d_{ij})^{-\alpha_L},       & \text{LoS}\\
\mu \beta_0(d_{ij})^{-\alpha_N},     & \text{NLoS}
\end{array} \right.
\end{equation}
where $\beta_0$ is the average channel power at a reference distance of 1 meter in the LoS state. $\mu$ is the signal attenuation factor due to the NLoS propagation. $\alpha_L$ and $\alpha_N$ are the average path loss exponents for LoS and NLoS states in time step $t$, respectively. $d_{ij}$ is the direct distance between the $i$-th IoTD and the $j$-th UAV.

\subsection{Data Collection Model}
In the proposed system, UAVs can either charge multiple IoTDs simultaneously or collect data from multiple IoTDs simultaneously. Hence, the whole process can be divided into two stages: the energy transfer stage and the data transmission stage.

\subsubsection{Energy Transfer Stage}Energy transfer technology is applied in the energy transfer stage, where the phased antenna array on UAVs would generate multiple beams to charge multiple IoTDs at the same time\cite{9806419}. By charging multiple devices simultaneously, it greatly improves the efficiency of energy transfer. 

In the system, the $j$-th UAV can transmit the energy to the $i$-th IoTD wirelessly via Radio Frequency (RF) signals with transmit power $P^{T}$. We assume the uplink power gain, downlink power gain, and the large-scale channel power gain are equal \cite{ 2020-aoi-eh}, which is described in Eq. \eqref{eq:prob_gain}. Hence, the power received at the $i$-th IoTD from the $j$-th UAV in time step $t$ is denoted as $P^{R}_{ij}[t] $ as follows:
\begin{equation}
\label{eq:eh_receive}
P^{R}_{ij}[t]=h_{ij}[t] P^{T}.  
\end{equation}

According to the Eq. \eqref{eq:prob_gain}, the expected receiving power is given as
\begin{equation}
\label{eq:expect_eh_receive}
P^{R}_{ij}[t]=p_{ij}^L[t] \beta_0(d_{ij})^{-\alpha_L} P^{T} + (1-p_{ij}^L[t]) \mu \beta_0(d_{ij})^{-\alpha_N} P^{T}.  
\end{equation}

We assume the hovering point of UAVs are set as the Cluster Heads (CHs) for a group of IoTDs, and IoTDs belonging to one CH would be collected by the specific UAV at the top of the CH \cite{aoi_cluster_head}. The proposed system adopts the linear energy transfer model, and the received energy of the $i$-th IoTD from the $j$-th UAV in time step $t$ can be given by

\begin{equation}
\label{eq:linear_eh_model}
E^{R}_{ij}[t]=\eta^{L} P^{R}_{ij}[t] T_{ij}^{E}[t]
\end{equation}
where $\eta^{L}$ is the constant attenuation parameter in the energy transfer model. $T_{ij}^{E}[t]$ is the energy transfer time of the $i$-th IoTD from the $j$-th UAV in time step $t$. 

\subsubsection{Data Transmission Stage} We introduce Orthogonal Frequency Division Multiplexing (OFDM) technology for the UAV to collect data from multiple IoTDs within its communication range simultaneously. The uplink data rate of the $i$-th IoTD to the $j$-th UAV in time step $t$ can be given by
\begin{equation}
\label{eq:jiang3}
R_{ij}[t]=B\log_2\left(1+\frac{h_{ij}[t] E^{R}_{ij}[t]}{\sigma^2 T^{C}_{ij}[t]}\right)
\end{equation}
where $B$ is the bandwidth, $\sigma^2$ is Gaussian white noise power, and $T^{C}_{ij}[t]$ is the data collection time from $i$-th IoTD to the $j$-th UAV in time step $t$. 

To ensure that IoTDs can successfully upload their data to the UAVs, one has
\begin{equation}
\label{eq:c_eh}
T^{C}_{ij}[t]R_{ij}[t] \geq D_i, \forall j\in \mathcal{M}
\end{equation}
where $D_i$ is the collected data from the $i$-th IoTD. 

\subsection{Flight Model}

In the proposed system, each UAV flies in a straight line from
one hovering point to another. The $j$-th UAV takes off from the
data center at a fixed location $r_j[0]=(0,0,H_z)$, flies to each CH as a designated hovering point, and hovers above it to collect data from all IoTDs within the corresponding cluster. The $j$-th
UAV completes the data collection task according to a predetermined flight trajectory, which consists of several hovering points, and returns to the same data center after one flight cycle. Hence, we have
\begin{equation}
\label{eq:c_return}
r_j[s_j]=r_j[0],\forall j \in\mathcal{M}
\end{equation}
where $r_j[t]$ denotes the position of the $j$-th UAV at the $t$-th hovering point,
%the $t$-th time step of the $j$-th UAV, 
$t \in\mathcal{T}_j=\left\{ 1,2,...,s_j \right\}$, and the trajectory of the $j$-th UAV has $s_j$ hovering point. 

We assume that each UAV flies at a constant speed $v_j$.
Since each UAV flies from one hovering point to another in a straight line, the flight time of the $j$-th UAV can be expressed by:
\begin{equation}
\label{eq:uav_flight_time}
T_{j}^{F}=\sum_{t=1}^{s_j}\frac{\left\|r_j[t]-r_j[t-1]\right\|_2}{v_j},  \forall j \in \mathcal{M}
\end{equation}
where $\left\|r_j[t]-r_j[t-1]\right\|_2$ is the Euclidean distance between hovering point $r_j[t]$ and $r_j[t-1]$.

\subsection{Energy Consumption Model}

Assuming that the flight power $P_{j}^F$ of the $j$-th UAV is constant, the flight energy consumption of the $j$-th UAV can be expressed as follows.
\begin{equation}
\label{eq:uav_flight_energy}
E_j^F= P_{j}^{F} {T_j^F}
\end{equation}
where $T_j^F$ is the total flight time of the $j$-th UAV.

Similarly, assuming that the hovering power $P_{j}^H$ of the $j$-th UAV is constant, the hovering energy consumption of the $j$-th UAV can be expressed as follows.
\begin{equation}
	\label{eq:uav_hover_energy}
	E_{j}^{H}=\sum_{t=1}^{T_j}({P^{H}T^E_j[t]}+P^{H} T^C_j[t])
\end{equation}
where $T^E_j[t]=\max_{i\in \mathcal{C}_k}T^{E}_{ij}$ is the maximum power transfer time for IoTDs in the $k$-th cluster, and $\mathcal{C}_k$ is the $k$-th cluster of IoTDs. $T^{C}_j[t]=\max_{i\in \mathcal{C}_k}T^{C}_{ij}$ is the maximum data collection time for IoTDs in the $k$-th cluster, and each IoTD applies the OFDM channel.

When the $j$-th UAV hovers at the top of one CH, it transfers its power to charge IoTDs first and then collects data from them. Hence, the power transfer energy of the $j$-th UAV is
 \begin{equation}
 \label{eq:jiang4}
E_j^T=\sum_{t=1}^{T_j}P^{T}T^E_j[t]
\end{equation}

Therefore, the total energy consumption of the $j$-th UAV can be given as
\begin{equation}
\label{eq:total_energy}
E_j=E_j^F+E_{j}^{H}+E_j^T, \forall j\in\mathcal{M}.
\end{equation}

Due to the limited data and energy capacity of UAVs, it is required that the total energy consumption of the UAV does not exceed its energy capacity, and all collected data does not exceed the storage capacity. Therefore, these inequalities need to be satisfied
\begin{equation}
\label{eq:c_battery}
E_j\leq E_{\max}, \forall j\in\mathcal{M},
\end{equation}
\begin{equation}
\label{eq:c_capacity}
\sum_{i=1}^N \sum_{t=1}^{s_j}a_{ij}[t]D_{i}\leq C_{\max}, \forall j\in\mathcal{M}.
\end{equation}

\subsection{Problem Formulation}
We aim to minimize the energy consumption of all the UAVs by jointly optimizing the number of UAVs, the trajectories of UAVs, user association, the energy transfer time, and the data collection time.
The optimization problem can be mathematically formulated by
\begin{equation}
\label{eq:problem}
\begin{aligned}
	P0:\min_{\mathcal{A},m,\mathcal{R}, \mathcal{T}^{E}, \mathcal{T}^{C}}\sum_{j=1}^m E_j \ \ \ \ \ \ \ \  \\
	s.t.~ \eqref{eq:c_uav_collect}, \eqref{eq:c_iot_transfer}, \eqref{eq:c_eh}, \eqref{eq:c_return}, \eqref{eq:c_battery},\eqref{eq:c_capacity}
\end{aligned}
\end{equation}
where $\mathcal{A}=\left\{a_{ij}[t], \forall i \in \mathcal{N}, j \in \mathcal{M},  t \in \mathcal{T}_j \right\}$ represents the association between UAVs and IoTDs. $m$ is the optimal number of UAVs for data collection. $\mathcal{R}=\left\{r_j[t], \forall j \in \mathcal{M},  t \in \mathcal{T}_j \right\}$ represents the visiting order of hovering points for UAVs. $\mathcal{T}^{E}=\left\{T_{ij}^{E}, \forall i \in \mathcal{N}, j \in \mathcal{M}, t \in \mathcal{T}_j \right\}$ represents the set of energy transfer time and $\mathcal{T}^{C}=\left\{T_{ij}^{C}, \forall i \in \mathcal{N}, j \in \mathcal{M}, t \in \mathcal{T}_j \right\}$ represents the set of data collection time. Problem $P0$ is a Mixed Integer Non-Linear Program (MINLP), where $E^F_j$ is affected by the visiting order of hovering point, $E^H_j$ and $E^T_j$ are influenced by energy transfer time and data collection time.

\subsection{Problem Decomposition}

In problem $P0$, there are an overwhelming number of terms that need to be optimized; however, it can be separated into different sub-problems which can be solved one at a time. First, we assume the number of hovering points is $K$, and we denote $\mathcal{H}=\left\{H_1, ..., H_K\right\}$ as the CHs, where $H_i$ denotes the coordinates of the $i$-th hovering point. \textcolor{black}{Then, we use the large-scale path-loss fuzzy c-means clustering algorithm (LS-FCM)\cite{2019jiang} to optimize the hovering points of UAVs and the association $\mathcal{A}$, which is based on the large-scale path-loss factors. This indicates that the proposed framework adopts a hierarchical solving strategy, where the hovering points are optimized first, followed by path planning and resource scheduling.} 
Then, we decompose the problem $P0$ into two sub-problems: energy transfer and data collection time optimization $P1$, and path optimization $P2$.

Since the energy transfer time and the data collection time of each IoTD are independent of the trajectories of UAVs, Problem $P1$ can be described as follows:
\begin{equation}
\label{eq:jiang5}
\begin{aligned}
P1:\min_{\mathcal{T}^{E}, \mathcal{T}^{C}}\sum_{j=1}^{m} (E^H_j+E^T_j) \\
s.t.~ \eqref{eq:c_eh}.\ \ \ \ \ \ \ \ \ \ \ \ 
\end{aligned}
\end{equation}

Meanwhile, Problem $P2$ can be described as follows:
\begin{equation}
\label{eq:jiang6}
\begin{aligned}
P2:\min_{m,\mathcal{R}}\sum_{j=1}^m E^F_j\ \ \ \\
s.t.~       
 \eqref{eq:c_return}, 
\eqref{eq:c_battery},\eqref{eq:c_capacity}.
\end{aligned}
\end{equation}

Problem $P2$ can be formulated as a well-known combinatorial optimization problem called the Multiple Traveling Salesman Problem (MTSP), which is known to be NP-hard. Moreover, each UAV has its limitations, and the number of UAVs is also required to be optimized. Hence, it is difficult to find the optimal solution by known methods.

\section{SLM-Based Solution Framework}
\label{sec:slm}
\begin{figure}[htbp]
	\centering
	\includegraphics[width=7cm]{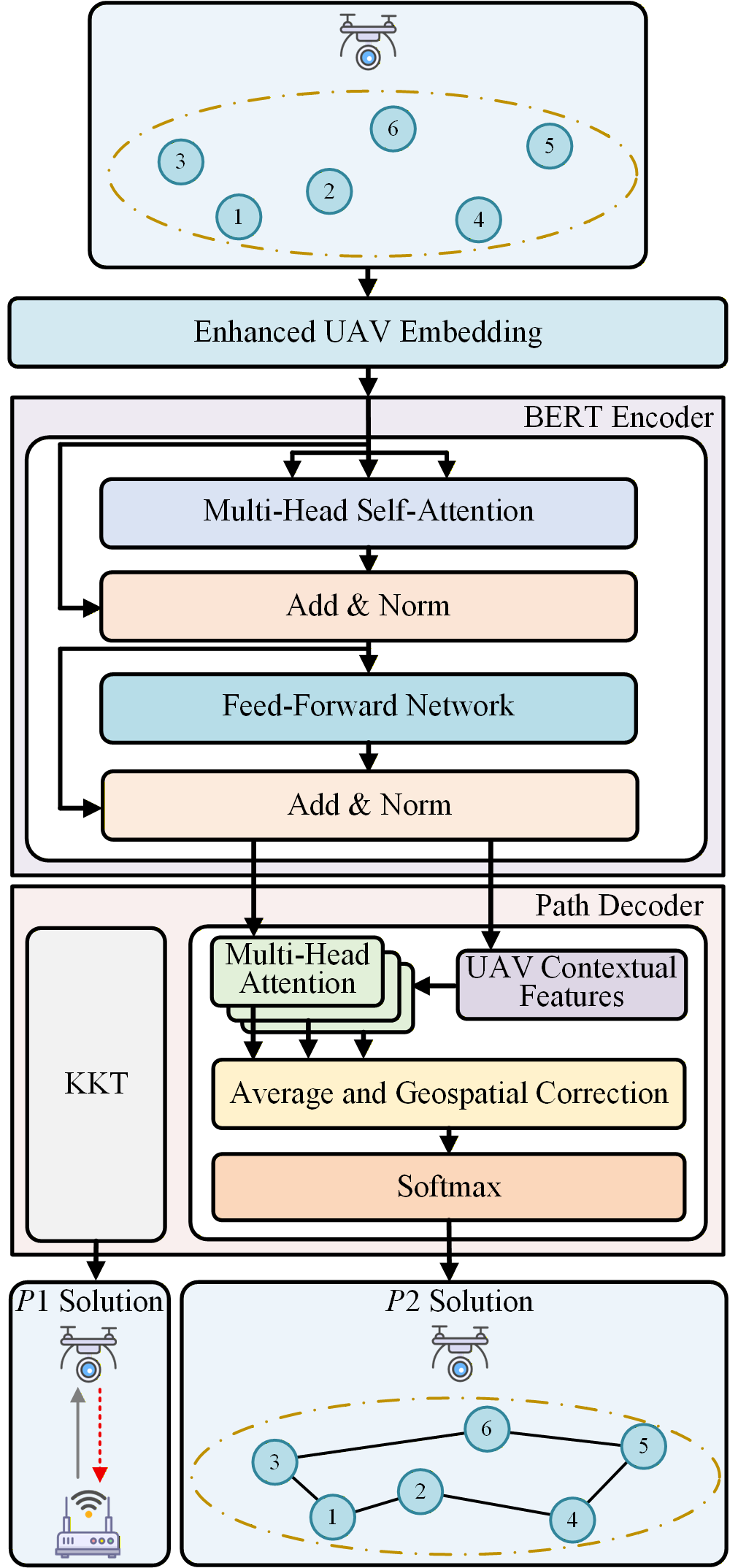}
	\caption{\textcolor{black}{The SLM-Based Solution Framework.}}
	\label{fig:bert}
\end{figure}

By combining SLMs with a path decoder, rapid UAV path planning and time allocation can be achieved at the network edge with low latency, low energy consumption, and high determinism, thus improving the performance-efficiency tradeoff in resource-constrained scenarios. Therefore, we propose an SLM scheme based on a lightweight BERT model \cite{10423879}. Since the original BERT architecture is designed for language data, we adapt its embedding layer, path decoder, and training strategy to suit energy-efficient UAV-enabled WPT systems. The proposed SLM is illustrated in Fig. \ref{fig:bert}.

\subsection{Enhanced UAV Embedding}

\subsubsection{Positional Feature Enhancement} 
In the BERT-based model, each node on the trajectory is typically represented by its three-dimensional coordinates $(x_i, y_i, z_i)$. While this representation encodes the absolute spatial position of each node, it lacks explicit geometric cues such as orientation and angular relationships in 3D space. Consequently, the model has to rely solely on Euclidean distances to implicitly infer the relative spatial structures among nodes, which increases the difficulty of geometric reasoning and trajectory reconstruction. 

To address this limitation, we introduce directional features $(\theta_i, \phi_i)$ into the input, representing the \textbf{azimuth} and \textbf{elevation} angles of each node with respect to the global coordinate system:
\begin{equation}
	\begin{aligned}
		\theta_i &= \text{atan2}(y_i, x_i) \\		
	\end{aligned}
\end{equation}
\begin{equation}
	\begin{aligned}
\phi_i &= \text{atan2}\left(z_i, \sqrt{x_i^2 + y_i^2}\right)
	\end{aligned}
\end{equation}
where $\theta_i$ denotes the azimuth angle of node $i$ in the $x$–$y$ plane, while $\phi_i$ represents its elevation angle relative to the horizontal plane. These angular descriptors provide the model with richer geometric context and facilitate more accurate spatial reasoning in trajectory modeling.

The enhanced node feature vector is thus obtained as:
\begin{equation}
\begin{aligned}
f_i = [x_i, y_i, \theta_i, \phi_i]
\end{aligned}
\end{equation}
and is then mapped through a linear embedding layer to an initial representation compatible with the BERT model:
\begin{equation}
\begin{aligned}
\mathbf{H}_i^{(0)} = \mathbf{W}^E f_i
\end{aligned}
\end{equation}
where $\mathbf{W}^E$ denotes a learnable embedding matrix.

This feature enhancement allows the encoder to not only capture the spatial distribution of nodes but also recognize their geometric orientation relative to the origin. Consequently, the model's understanding of path geometry is improved, enabling the self-attention mechanism to more easily learn the relative positional patterns of all hovering points.

\subsubsection{Geometric Feature Enhancement}
The MTSP inherently exhibits rotational and reflective symmetry in geometric space; that is, the optimal length of any path remains unchanged when the entire set of points undergoes planar rotation or mirroring. This property provides an effective means for data augmentation. Specifically, we employ the square symmetry group (i.e., the dihedral group $D_4$) \cite{sharma2020dihedral} to apply the following eight geometric transformations to the original set of coordinates:
\begin{equation}
\begin{aligned}
	g_1 &: (x, y) \mapsto (x, y), && \text{Identity (no transformation)}\\
	g_2 &: (x, y) \mapsto (-x, y), && \text{Reflection over the $y$-axis}\\
	g_3 &: (x, y) \mapsto (x, -y), && \text{Reflection over the $x$-axis}\\
	g_4 &: (x, y) \mapsto (-x, -y), && \text{Rotation by $180^\circ$}\\
	g_5 &: (x, y) \mapsto (y, x), && \text{Reflection over the line $y = x$}\\
	g_6 &: (x, y) \mapsto (-y, x), && \text{Rotation by $90^\circ$}\\
	g_7 &: (x, y) \mapsto (y, -x), && \text{Rotation by $270^\circ$}\\
	g_8 &: (x, y) \mapsto (-y, -x), && \text{Reflection over the line $y = -x$}.
\end{aligned}
\end{equation}

Based on the above geometric transformations, each topological graph instance can be converted into eight equivalent instances. All instance features are then used as the input to the initial embedding layer. \textcolor{black}{Theoretically, these symmetry-enhanced augmentations preserve the optimal path and underlying policy landscape, thereby maintaining policy consistency while increasing training diversity and improving generalization \cite{bronstein2021geometric}.}
Through positional and geometric feature enhancement, the expressive capacity of node embeddings is significantly enriched, improving BERT’s ability to capture the geometric characteristics of paths. % Through positional and geometric feature enhancement, the expressive capacity of node embeddings is significantly enriched, thereby improving BERT’s ability to capture the geometric characteristics of paths.

\subsection{BERT Encoder}

 The BERT model consists of $L$ encoder layers. In the $\ell$-th encoder layer, the Multi-Head Self-Attention (MHSA) is first computed as \cite{devlin2019bert}:
\begin{equation}
\begin{aligned}
\text{Attention}(\mathbf{Q}, \mathbf{K}, \mathbf{V}) = \text{Softmax} \left( \frac{\mathbf{Q}\mathbf{K}^T}{\sqrt{d_k}} \right) \mathbf{V}
\end{aligned}
\end{equation}
where the query $\mathbf{Q} = \mathbf{H}^{(\ell-1)}\mathbf{W}^Q$, key $\mathbf{K} = \mathbf{H}^{(\ell-1)}\mathbf{W}^K$, and value $\mathbf{V} = \mathbf{H}^{(\ell-1)}\mathbf{W}^V$, with $\mathbf{W}^Q, \mathbf{W}^K, \mathbf{W}^V$ being learnable weight matrices. For $H$ attention heads, the outputs are concatenated and projected as:
\begin{equation}
\begin{aligned}
\text{MHSA}(\mathbf{H}^{(\ell-1)}) = [\text{head}_1; \dots; \text{head}_H] \mathbf{W}^O
\end{aligned}
\end{equation}
where $\mathbf{W}^O$ is a learnable matrix, and $\text{head}_h = \text{Attention}(\mathbf{Q}_h, \mathbf{K}_h, \mathbf{V}_h)$.

Next, residual connection (Add) and layer normalization (Norm) are applied:
\begin{equation}
\begin{aligned}
\mathbf{Z}^{\ell} = \text{Norm}\left(\mathbf{H}^{(\ell-1)} + \text{MHSA}(\mathbf{H}^{(\ell-1)}) \right).
\end{aligned}
\end{equation}

The output is then passed through a Feed-Forward Network (FFN), followed by another residual connection and LayerNorm:
\begin{equation}
\begin{aligned}
\mathbf{H}^{\ell} = \text{Norm}(\mathbf{Z}^{\ell} + \text{FFN}(\mathbf{Z}^{\ell})).
\end{aligned}
\end{equation}

By stacking $L$ layers of such encoders, BERT can obtain bidirectional contextual feature representations, ultimately producing $\mathbf{H}^{L}$, where each row corresponds to the contextual representation of a node in the input sequence.

\subsection{Enhanced UAV Contextual Features}
For high-quality path planning, it is crucial to construct a query vector $\mathbf{q}[t]$ based on the encoder output $\mathbf{H}^{L}$, which fully captures the current path state and guides the selection of the next node. Unlike approaches that rely solely on features derived from the “current node and global node mean” \cite{khalil2017learning}, we explicitly incorporate four types of feature information—starting node, previous node, global structure, and path structure—to form an enhanced UAV contextual feature representation.
\subsubsection{Starting Node Feature Vector}
$\mathbf{h}_{\pi_{[1]}}$ represents the feature vector of the starting node (typically the data center). Incorporating this ensures that the SLM model is aware of the overall starting constraint during path planning, thereby helping maintain route closure and overall feasibility when generating paths.
\subsubsection{Previous Node Feature Vector}
$\mathbf{h}_{\pi_{[t-1]}}$ represents the feature vector of the most recently visited node along the current path. It provides local positional information, serving as the most direct spatial reference for the decoder when selecting the next node to visit.

\subsubsection{Global Feature Vector}
 $\mathbf{h}_g = \frac{1}{N}\sum_{i=1}^{N} \mathbf{h}_i$ represents the global structure of all nodes. This pooled representation captures the overall layout, providing a macroscopic perspective for decision-making and preventing the path from relying solely on local information, which could lead to greedy choices.

\subsubsection{Path Feature Vector}
 $\mathbf{h}_\pi = \sum_{s=1}^{t-1} \mathbf{h}_{\pi_{[s]}}$ encodes the set of nodes already visited along the path, allowing the model to explicitly consider the coverage and cumulative cost of the historical trajectory. This enhances global consistency and robustness in path planning.

By aggregating the above feature vectors, the enhanced UAV contextual feature query is formulated as:
\begin{equation}
\begin{aligned}
\mathbf{q}[t] = \frac{1}{t-1} \left(\mathbf{h}_\pi\right) + \mathbf{h}_g + \mathbf{h}_{\pi_{[t-1]}} + \mathbf{h}_{\pi_{[1]}}.
\end{aligned}
\end{equation}

As a result, the query vector $\mathbf{q}[t]$ semantically incorporates global, local, and historical contextual information, significantly enhancing the decision-making capability of the subsequent path decoder. Intuitively, this construction prevents the SLM model from making decisions based solely on the current node and global mean, thereby better balancing overall path feasibility and local optimality, and effectively improving the quality and generalization of path planning.

\subsection{Path Decoder}

We design a geometry-aware path decoder to sequentially generate UAV paths. In the path decoder, the query vector $\mathbf{q}[t]$ interacts with the contextual features of candidate nodes to compute the selection probability for each node, enabling step-wise path generation.

During the decoding phase, the query vector $\mathbf{q}[t]$ and the encoding $\mathbf{h}_j$ of the $j$-th candidate node are first projected via separate multi-head linear projections. Scaled dot-product attention is then applied to compute multi-head interaction scores. Averaging over $H$ attention heads yields the attention score for the $j$-th candidate node:
\begin{equation}
\begin{aligned}
\text{s}_{j} [t] = \frac{1}{H} \sum_{h=1}^{H} \frac{(\mathbf{q}[t] \mathbf{W}_h^Q)(\mathbf{h}_j \mathbf{W}_h^K)^{T}}{\sqrt{d_k}}
\end{aligned}
\end{equation}
where $\mathbf{W}_h^Q$ and $\mathbf{W}_h^K$ denote the projection matrices for the $h$-th attention head, and $d_k$ is the hidden dimension of each head.

To integrate data-driven attention scores with problem-specific geometric priors, we introduce a proximity heuristic. The basic idea is that UAV paths tend to connect spatially nearby nodes, so scoring should explicitly favor closer candidates. Accordingly, the adjusted attention score for the $j$-th candidate node is defined as:
\begin{equation}
\begin{aligned}
\text{S}_{j} = \text{s}_{j} [t] - \text{Geo}_j
\end{aligned}
\end{equation}
where $\text{Geo}_j$ represents the cost value from the current node to candidate node $j$, typically taken as the Euclidean distance, but it can also be extended to include energy consumption, flight time, or other task-related cost functions. By adjusting the attention scores in this manner, even if the data-driven attention mechanism tends to favor distant nodes, the proximity heuristic corrects this bias, ensuring that the generated path is geometrically reasonable.

After obtaining the adjusted scores, we calculate the selection probability $p_j[t]$, which represents the probability that the current UAV selects the $j$-th node as the next hovering point to collect data in the $t$-th time step:
\begin{equation}
	\begin{aligned}
		p_j [t] = \text{Softmax} (C \cdot \tanh(\text{S}_{j}))\mathbf{M}_j
	\end{aligned}
\end{equation}
where $C$ is a scaling factor and $\mathbf{M}$ is the masking. Initially, we set $M = [1, ..., 1]$ at the beginning of path planning. The length of $\mathbf{M}$ is $N$. When $\mathbf{M}_i = 1$ means the UAV can select the $i$-th node. There are three cases when $\mathbf{M}_i = 0$: (1) The $i$-th node has already been visited; (2) the UAV cannot fly to the position of the $i$-th node due to insufficient battery power; and (3) the UAV cannot collect data from the $i$-th node due to limited storage capacity. At this point, Problem $P$2 has been solved.

Then, since Problem  $P1$ is a convex optimization problem after the trajectory and associations are fixed. The optimal data collection time $\mathcal{T}^{C}$ and energy transfer time $\mathcal{T}^{E}$ are achieved when Eq. (\ref{eq:c_eh})  is satisfied with equality as follows:
\begin{equation}
	\label{eq:t_cd_convex}
	{T^{C}_{ij}}[t]^{*}B\log_2 \left(1+\frac{h_{ij}[t] {E^{R}_{ij}}[t]^{*}}{\sigma^2 {T^{C}_{ij}}[t]^{*}}\right)=D_{i}, \forall j \in \mathcal{M}, \forall i \in \mathcal{N}.
\end{equation}

The convex optimization problem can be solved by applying the Karush–Kuhn–Tucker (KKT) conditions \cite{2012-kkt}, and the corresponding optimal solutions $\left\{ {T^{C}_{ij}}[t]^{*}, {T^{E}_{ij}}[t]^{*} \right\}$ are then derived following the analytical procedure in \cite{2020-aoi-eh}.
Hence, we derive the optimal energy transfer time and data collection time using the KKT conditions, based on which the Problem $P$1 is solved.

\subsection{Complexity Analysis}

\textcolor{black}{Let $n$ denote the number of hovering points, $d$ the hidden dimension, and $L$ the number of encoder layers. The computational complexity of the proposed SLM is mainly dominated by the BERT encoder and the path decoder. Specifically, the complexity of each encoder layer is $O(n^{2}d + nd^{2})$, and thus the total complexity of the encoder is $O\!\left(L(n^{2}d + nd^{2})\right)$. The path decoder needs to score all candidate nodes while sequentially generating a complete path, resulting in a complexity of $O(n^{2}d)$. By contrast, the enhanced embedding and contextual feature construction only involve linear projection and aggregation, with complexity $O(nd)$, which is negligible compared with the encoder and decoder. Therefore, the overall single-view inference complexity of the SLM can be expressed as $O\!\left(L(n^{2}d + nd^{2}) + n^{2}d\right)$.}
%This indicates that the computational complexity is mainly dominated by the quadratic term with respect to the number of hovering points $n$.

\subsection{Ensemble Inference}

During the inference phase, we leverage the symmetry of the $D_4$ group to further enhance solution quality and robustness. For each geometrically equivalent transformation $g \in \mathcal{G}$, a new instance is generated under the transformation and decoded independently to obtain the path $\pi_g$. The path $\pi_g$ is then mapped back to the original coordinate system via the inverse transformation $g^{-1}(\pi_g)$. Finally, multiple solutions obtained under different equivalent coordinate systems are compared, and the optimal one is selected as the final output:
\begin{equation}
\begin{aligned}
\pi^* = \arg \min_{g \in \mathcal{G}} L(g^{-1}(\pi_g))
\end{aligned}
\end{equation}
where $L$ denotes the objective function, which in this work corresponds to Problem $P2$. Compared to single-view inference, this approach significantly reduces the risk of suboptimal solutions caused by model bias or numerical perturbations, thereby improving overall stability and reliability. Moreover, ensemble inference ensures consistent model performance across different coordinate orientations, substantially enhancing cross-distribution generalization and preventing the model from being trapped in local optima due to attention biases under a specific geometric arrangement.

\subsection{Reinforcement Path Optimization}

\subsubsection{Policy Gradient Update}
During the training phase, we adopt a policy gradient-based reinforcement learning framework to optimize the path quality generated by the path decoder in an end-to-end manner. Specifically, the model parameters $\theta$ are optimized to minimize the expected path length.
To further improve training efficiency and leverage the geometric symmetry inherent in path planning, for a given instance, paths are decoded simultaneously from multiple different starting points, resulting in a set of equivalent complete paths. Let the path set obtained for instance $\mathcal{I}_i$ from $N$ different starting points be $\{ \tau_{ij} \}_{j=1}^{N}$. The returns of all paths are then used to update the policy gradient:
\begin{equation}
	\begin{aligned}
		\nabla_{\theta} J(\theta) &\approx \frac{1}{BN} \sum_{i=1}^{B} \sum_{j=1}^{N} \hat{R}(\tau_{ij}) \nabla_{\theta} \log p_{\theta}(\tau_{ij} | \mathcal{I}_i)
	\end{aligned}
\end{equation}
where $B$ denotes the batch size, $\tau_{ij}$ represents the $j$-th sampled path sequence for instance $\mathcal{I}_i$, and $\hat{R}(\tau_{ij})$ is the normalized reward. Moreover, since the path length distributions can vary significantly across different instances, directly using absolute path length as the reward may lead to large variations in cross-instance gradient updates. To address this, rewards are normalized using the mean and standard deviation within each instance:
\begin{equation}
	\begin{aligned}
	\hat{R}(\tau_{ij}) = \frac{R(\tau_{ij}) - \mu(\mathcal{I}_i)}{\sigma(\mathcal{I}_i)}
	\end{aligned}
\end{equation}
where $\mu(\mathcal{I}_i)$ and $\sigma(\mathcal{I}_i)$ denote the mean and standard deviation of the rewards across all equivalent paths for instance $\mathcal{I}_i$. \textcolor{black}{Theoretically, reward normalization can effectively reduce the variance of policy gradient estimation across samples with different reward scales, thereby mitigating optimization oscillation and training instability caused by inconsistent return distributions. Furthermore, by mapping the rewards of different instances into a comparable numerical range, this mechanism provides smoother and more consistent learning signals for policy updates, thus improving training stability and accelerating convergence\cite{schulman2017proximal}.} % This normalization stretches the reward distribution of each instance to zero mean and unit variance, effectively reducing scale differences across instances and significantly improving training stability and convergence speed.

Additionally, during training, a subset of paths from the set $\{ \tau_{ij} \}_{j=1}^{N}$ is randomly sampled and subjected to $D_4$ group-equivalent transformations, generating samples that are geometrically equivalent to the original problem but have different distributions. This not only significantly increases the diversity of the training data but also prevents the SLM model from overfitting to specific coordinate systems or regional distributions, enabling it to learn geometry-invariant features.
\subsubsection{Reward Function Design}

During training, the optimization objective of the model parameters $\theta$ is equivalent to maximizing the expected return of the negative path length. The reward function is thus defined as the negative of the path length. In addition, a penalty term is introduced based on the path length to assign negative rewards to solutions that violate constraints:
\begin{equation}
\begin{aligned}
R(\tau) = -L_{\theta}(\tau) - \lambda \cdot (\Phi_E(\tau)+\Phi_C(\tau))
\end{aligned}
\end{equation}
where $L_{\theta}(\tau)$ denotes the distance function, $\lambda > 0$ is the penalty coefficient, $\Phi_{E}(\tau)$ quantifies the degree of violation of the energy constraint for path $\tau$, and $\Phi_{C}(\tau)$ quantifies the degree of violation of the storage constraint for path $\tau$.

For the energy constraint, it is defined as
\begin{equation}
\begin{aligned}
\Phi_{\text{E}}(\tau) = \sum_{k=1}^{K} \max(0, E_k(\tau) - E_k^{\max})
\end{aligned}
\end{equation}
where $E_k(\tau)$ denotes the energy consumption of UAV $k$ along path $\tau$, and $E_k^{\max}$ is its maximum battery capacity.

Similarly, for the storage constraint, it is formulated as
\begin{equation}
	\begin{aligned}
		\Phi_{\text{C}}(\tau) = \sum_{k=1}^{K} \max(0, C_k(\tau) - C_k^{\max})
	\end{aligned}
\end{equation}
where $C_k(\tau)$ represents the storage requirement of UAV $k$ along path $\tau$, and $C_k^{\max}$ denotes its maximum storage capacity.

\section{Solution Framework from LLM to Agentic AI}
\label{sec:llm}

\subsection{LLM-Based Solution Framework}
Multimodal Large Language Models (MLLMs) demonstrate unique advantages in energy-efficient UAV-enabled WPT systems. Traditional methods rely on numerical coordinates and distance matrices to optimize paths, whereas MLLMs can reason directly from multimodal inputs such as images and text, leveraging visualized point distribution information to generate feasible paths. This capability emulates human spatial intuition and visual reasoning, avoiding complex numerical computations. Moreover, MLLMs exhibit strong generalization in zero-shot and few-shot scenarios, enabling rapid generation of near-optimal solutions without task-specific training data \cite{11175646}.

\begin{figure}[htbp]
	\centering
	\includegraphics[width=7cm]{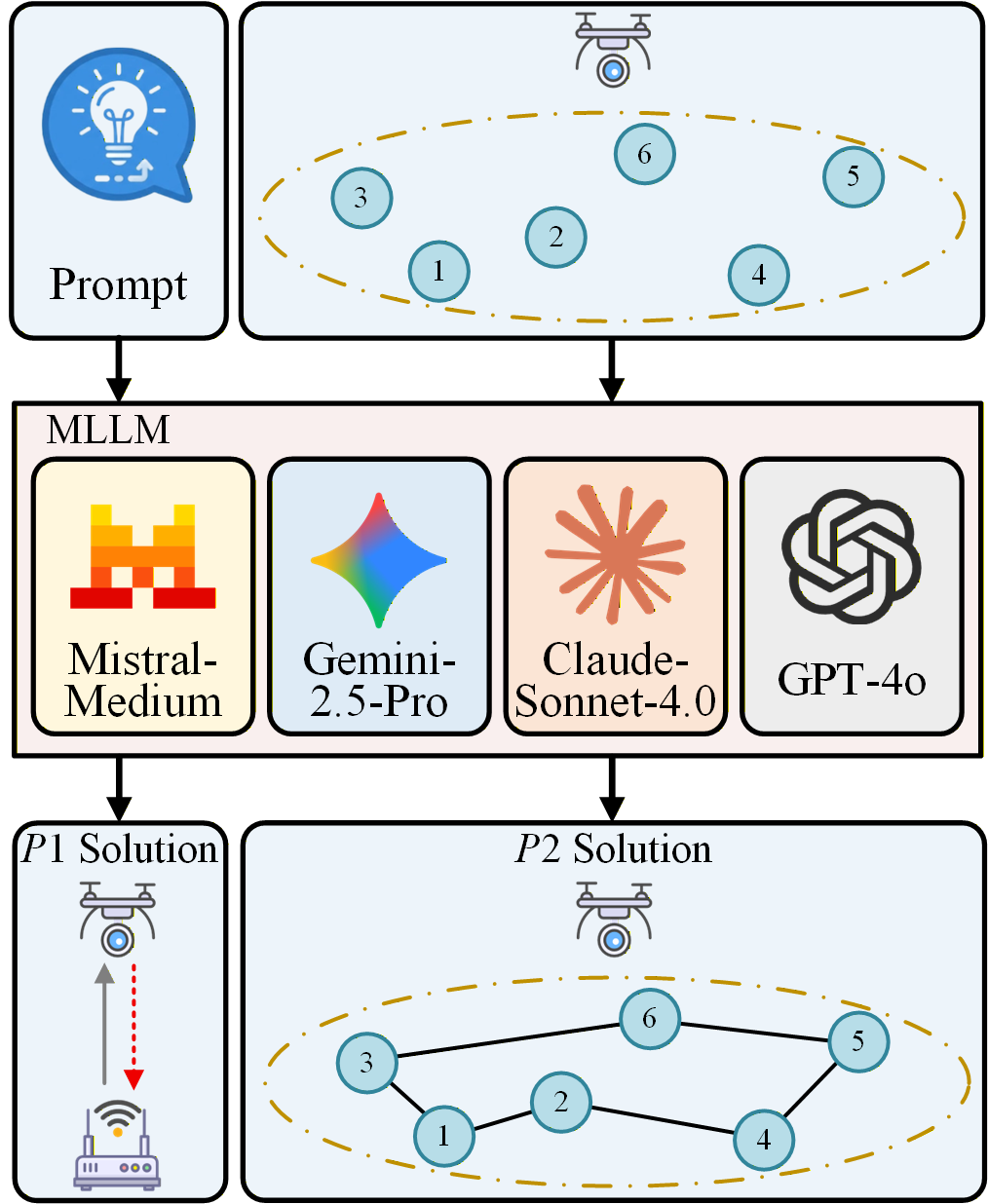}
	\caption{\textcolor{black}{The MLLM-Based Solution Framework.}}
	\label{fig:MLLM}
\end{figure}

We design an MLLM-based solution framework that employs visual reasoning and chain-of-thought reasoning to simultaneously solve the following two problems: (i) optimization of energy transfer and data collection time, and (ii) optimization of UAV flight path.
First, structured prompts are constructed to explicitly inject the system model, parameters, and constraints into the MLLM, together with the visualized distribution of UAV hovering points. By reasoning directly over these multimodal representations, the MLLM can generate a feasible flight trajectory that attempts to satisfy the given constraints (defined as Problem $P$2).
Subsequently, continuous variable optimization is performed along this trajectory. Although most MLLMs cannot explicitly derive the KKT conditions, they are capable of generating executable Python code for solving them. By invoking an external code interpreter to run this generated code, we further obtain the optimal energy transfer time and data collection time, corresponding to Problem $P$1. The proposed MLLM-based solution framework  is illustrated in Fig. \ref{fig:MLLM}.

\subsection{Agentic AI-Based Solution Framework}
 Through multi-agent collaborative iterative optimization, MLLMs can effectively reduce path intersections, improve coverage completeness, and enhance solution robustness \cite{11370176}. These features make MLLMs a powerful tool for UAV path planning in complex environments with multiple constraints.

\begin{figure*}[htbp]
	\centering
	\includegraphics[width=17cm]{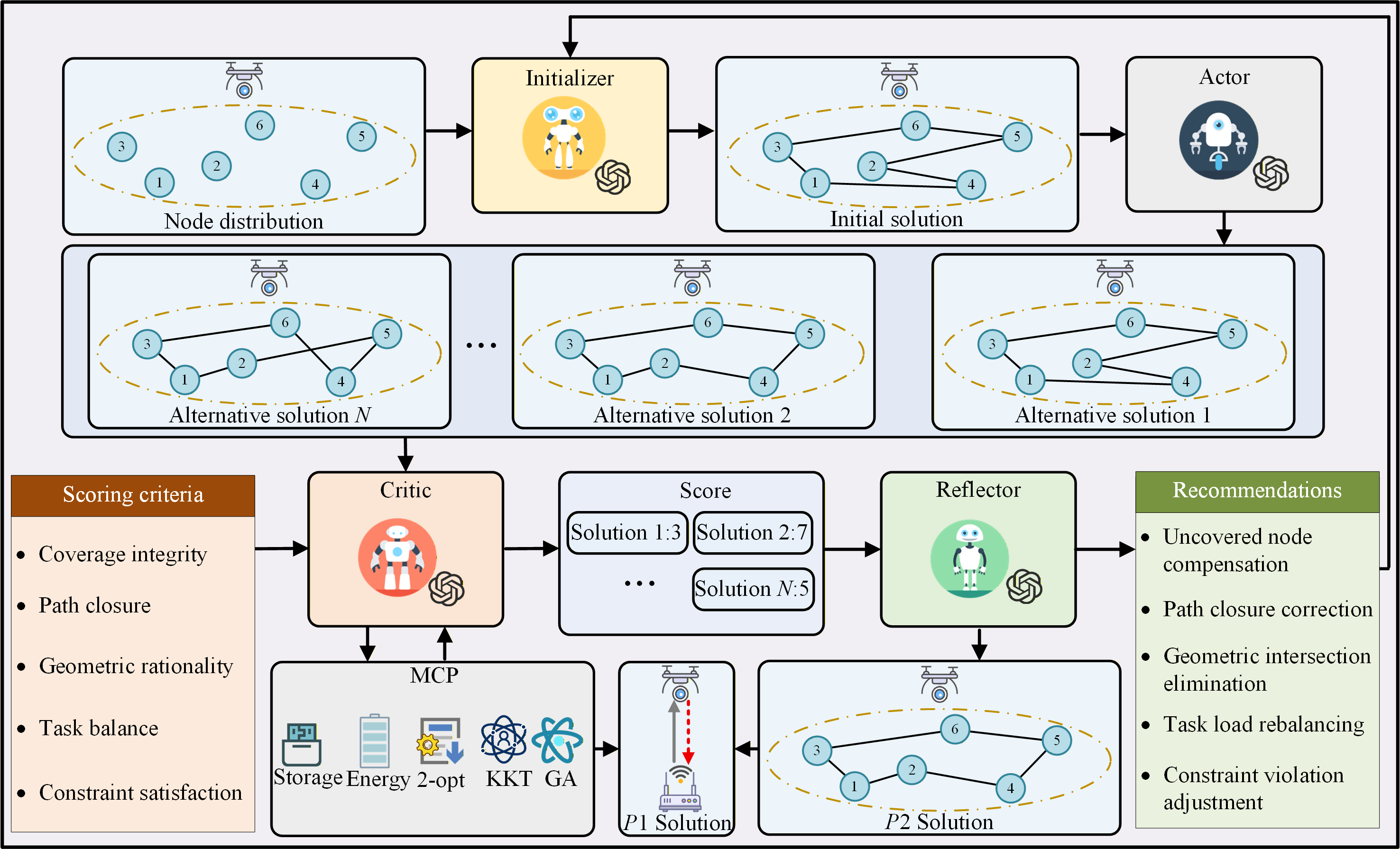}
	\caption{The Agentic AI-Based Solution Framework.}
	\label{fig:Agentic AI}
\end{figure*}

Based on these insights, we propose an Agentic AI system for solving UAV path planning problems. The core idea is to integrate MLLMs with the MCP to achieve intelligent path planning for UAV networks. The system comprises four types of agents: Initializer, Actor, Critic, and Reflector, which collaboratively form a closed-loop process of ``generation–optimization–evaluation–reflection" for UAV paths. The proposed Agentic AI system is illustrated in Fig. \ref{fig:Agentic AI}.

Specifically, the Initializer generates feasible initial path solutions, while the Actor produces diverse candidate paths via high-temperature sampling. The Critic evaluates the paths, invoking the MCP module to verify constraint satisfaction and provide scores for each candidate. The Reflector generates targeted improvement suggestions based on the scoring results, which are fed back to the Initializer, thereby guiding subsequent iterations in a more focused search direction. This framework significantly enhances system autonomy, interpretability, and constraint feasibility. The workflow of the Agentic AI system is summarized as follows:

\subsubsection{Initializer for Initial Solution Generation}

The goal of the Initializer is to construct a feasible initial path for the UAV system. The input consists of a map containing a data center and $N$ hovering nodes. This agent partitions the nodes such that each UAV has a closed path starting and ending at the data center, ensuring that every node is visited exactly once. Moreover, during iterative optimization, the Initializer leverages suggestions from the Reflector to make targeted adjustments, minimizing path redundancy and intersections to reduce the difficulty of subsequent optimization.

\subsubsection{Actor for Candidate Solution Optimization}

The Actor is responsible for improving upon the initial solutions generated by the Initializer. By employing high-temperature sampling (temperature = 0.7), the Actor can generate multiple candidate path planning solutions in a single iteration, thereby increasing the diversity of the solution space. Its optimization objectives include minimizing path length, balancing task load, and ensuring geometric feasibility of the paths.

\subsubsection{Critic for Candidate Solution Evaluation}

The Critic evaluates the quality of the solutions generated by the Actor. The performance of each candidate solution is decomposed into five dimensions: coverage completeness, path closure, geometric rationality, task balance, and constraint satisfaction. Constraint satisfaction is verified through the MCP protocol. Specifically, the MCP module invokes the system model code to rigorously check energy and storage constraints for each candidate path, returning the results to the Critic, which then computes individual scores and a weighted overall score for each solution.

\subsubsection{Reflector for Reflective Analysis and Improvement}

The Reflector is responsible for analyzing the scoring results provided by the Critic, selecting the current best path, diagnosing it, and generating targeted improvement suggestions. Its analysis scope includes filling in uncovered nodes, repairing path closure, reducing geometric intersections, rebalancing task loads, and adjusting for constraint violations. The current best path along with the improvement recommendations is then sent back to the Initializer, guiding the next iteration of the optimization process.

\subsubsection{Energy Transfer and Data Collection Time Optimization}
Upon obtaining the optimized path, and under prompt guidance, the MCP proceeds to invoke additional optimization tools, such as KKT-based or Genetic Algorithm (GA) solvers, to optimize the energy transfer time and data collection time, thereby solving Problem $P$1.

In the proposed Agentic AI framework, the interactions among agents are primarily guided by the visual reasoning capability of the MLLM, enabling multimodal perception and reflective collaboration for path generation and optimization. When the MLLM produces suboptimal planning results—such as path intersections or incomplete coverage—the Agentic AI system can invoke external local search algorithms (e.g., 2-opt or 3-opt) via the MCP for further refinement. This hybrid mechanism combines the global visual reasoning of the MLLM with the local optimization strength of heuristic search, significantly improving path optimality. 

%\textcolor{red}{To enhance safety and robustness, the framework also includes a verification layer that checks agent-generated actions against temporal-logic constraints and domain-specific safety rules. Potential threats, such as external interference or falsified state information, are mitigated through monitoring modules that validate trajectory feasibility and signal authenticity. Suboptimal or risky plans identified during MLLM reasoning can trigger external verification or local optimization routines, ensuring adherence to physical and operational constraints while maintaining adaptive decision-making.}

\section{Results and Discussion}
\label{sec:experiments}

\subsection{Simulation Settings}

Unless otherwise specified, the simulations are conducted in a $1000 ~\text{m} \times 1000 ~\text{m}$ square area to emulate a typical multi-UAV data collection scenario. In the area, 100 latent cluster centers are first randomly generated, and then 500 IoTDs are randomly distributed around these centers. The data size $D_i$ for each IoTD is randomly selected within the range of [0.2, 0.5] MB. The data center is located at the coordinate $(500, 500)$. All UAVs depart from the data center to perform their missions and return to the same location after completing a full flight cycle. Each UAV operates at a fixed altitude of $H_z = 20$ m. Other detailed parameters of the system model are summarized in Table \ref{tab:param}. \textcolor{black}{These settings are adopted as a baseline scenario for fair comparison. In particular, the LS-FCM algorithm is applied to determine the final UAV hovering points, the data center can be deployed at arbitrary locations, and the formulation can be further extended to heterogeneous UAV-capacity scenarios.}

In the SLM scheme, the DistilBERT model is adopted, consisting of 6 Transformer encoder layers, 12 self-attention heads, and a hidden dimension of 768, resulting in approximately 66 million parameters \cite{sanh2019distilbert}. To adapt DistilBERT to UAV path planning, we fine-tune the model by introducing an enhanced UAV embedding layer (positional and geometric-angle features), and designing an 8-head path decoder.

In the LLM and Agentic AI schemes, GPT-4o is utilized. Each agent adopts a temperature-controlled sampling mechanism (Temperature = 0.7) to balance exploration and stability during reasoning.

\begin{table}[h]
	\caption{Parameters of the System Model.}
	\centering
	\setlength{\tabcolsep}{3mm}
	\renewcommand\arraystretch{1.25}
	\begin{tabular}{|c|c|}\hline
		Parameter&{Value}\\\hline
		UAV data-storage capacity $C_{\max}$&50 MB\\
				UAV flight speed $v$ & 10 m/s\\
		Flight height $H_z$ & 20 m\\
		UAV battery capacity $E_{\max}$&2550 mAh\\
		Bandwidth $B$ & 2 MHz\\
		Transmit power $P^{T}$ & 10 W\\
		Flight power $P^{F}$ & 75 W\\
		Hover power $P^{H}$ & 50 W\\
		Signal attenuation factor $\mu$ & 0.8\\
		LoS probability factor $B_1,B_2,B_3,B_4$ & -0.45, 0.04, -0.63, 1.63\\
		Channel power gain $\beta_0$ & -30 dB\\
		Noise power $\sigma^{2}$ & -110 dB\\
		Attenuation parameter $\eta^{L}$ & 0.9\\
		Average path loss exponents $\alpha_L,\alpha_N$ & 2.5, 3.5\\
		\hline
	\end{tabular}
	
	\label{tab:param}
\end{table}

\textcolor{black}{The entire simulation framework is implemented in Python 3.8 using the PyTorch 1.13.0 deep learning library. All experiments are conducted on a workstation equipped with an Intel Xeon CPU and an NVIDIA A800 GPU with 80 GB of memory. For SLM training, LoRA-based fine-tuning is adopted, with the total number of training epochs set to 50 and the batch size set to 64. The reinforcement learning training dataset contains 1 million samples, and the minimum learning rate is set to $1 \times 10^{-5}$.}

\subsection{Experiment 1: Ablation Study of the SLM}

This experiment aims to systematically evaluate the incremental contribution of each structural component in the proposed SLM. The number of UAV hovering points is defined as 100, and a single UAV is assumed to visit all hovering points (without resource constraints).
To this end, five model variants are designed, where each variant progressively incorporates one additional module based on the previous configuration:

\begin{itemize}
	\item \textbf{SLM-Base:} The baseline model adopts a BERT with routing decoder \cite{2018attention}.
	\item \textbf{SLM-E:} The enhanced UAV embedding layer is introduced to encode geometric positional information.
	\item \textbf{SLM-C:} Building upon SLM-E, this variant further integrates an enhanced contextual feature query mechanism.
	\item \textbf{SLM-G:} The geometry-aware path decoder is incorporated.
	\item \textbf{SLM (Ours):} The complete version integrates all the above modules and the ensemble inference.
\end{itemize}

All model variants are trained under identical configurations using the reinforcement path optimization framework. 
Their performance is evaluated from three perspectives: the total energy cost  $E_{\text{tot}}$, the trajectory length $L_{\text{traj}}$, and the optimality gap $G_{\text{opt}}$. We define the optimality gap $G_{\text{opt}}$ as the relative sub-optimality with respect to the best-known solution:

\begin{equation}
\begin{aligned}
	G_{\text{opt}} &= \frac{L_{\text{traj}} - L_{\text{traj}}^{\star}}{L_{\text{traj}}^{\star}} \times 100\%
\end{aligned}
\end{equation}
where $L_{\text{traj}}^{\star}$ denotes the minimum trajectory length among all compared methods; $G_{\text{opt}}=0$ indicates that the current solution attains the best performance within the set of considered methods. The comparison results of SLM variants, together with the Optimal solution (OPT) and the conventional 2-opt baseline, are listed in Table \ref{tab:ablation}.

\begin{table}[h]
	\centering
	\caption{\textcolor{black}{Comparison of SLM Variants with OPT and 2-opt.}}
	\label{tab:ablation}
	\setlength{\tabcolsep}{5mm}
	\renewcommand\arraystretch{1.25}
	\resizebox{9cm}{!}{
		\color{black}
		\begin{tabular}{lccc}
			\toprule
			\textbf{Model Variant} & $E_{\text{tot}}$ (Wh) & $L_{\text{traj}}$ (m) & $G_{\text{opt}}$ (\%) \\
			\midrule
			OPT   & - & 7760.00 & 0  \\
			2-opt   & 38.98 & 9072.67 & 16.91  \\
			SLM-Base   & 36.72 & 7986.43 & 2.91 \\
			SLM-E      & 36.45 & 7856.95 & 1.25 \\
			SLM-C      & 36.38 & 7832.38 & 0.93 \\
			SLM-G      & 36.31 & 7798.46 & 0.49 \\
			SLM (Ours) & \textbf{36.22} & \textbf{7773.29} & \textbf{0.17} \\
			\bottomrule
	\end{tabular}}
\end{table}

The results show that each enhancement progressively improves performance. The full SLM achieves the minimum energy consumption and yields the shortest flight path. This improvement stems from three factors: (1) enhanced embeddings provide richer geometric cues, (2) contextual queries capture both global and historical dependencies, and (3) the path decoder grounded in geometric knowledge and ensemble inference introduces spatial symmetry awareness, thereby enhancing generalization to unseen UAV layouts.

\subsection{Experiment 2: Comparison with Other Path Planners}

To verify the superiority of the proposed SLM, we compare it with three lightweight path planning models: ATOM~\cite{10879146}, Pointerformer~\cite{jin2023pointerformer}, and AttRouting~\cite{2018attention}. The number of UAV hovering points is defined as 100, and a single UAV is assumed to visit all hovering points. All methods employ the same encoder architecture to ensure comparable parameter scales. The models are trained on identical datasets and evaluated in terms of the flight energy cost $E^{F}$, the standard deviation $StD$, and the training time $T_{\text{tra}}$ and the inference time $T_{\mathrm{inf}}$. The performances of all path planning models are listed in Table \ref{tab:smallpath}.

\begin{table}[h]
	\centering
	\caption{\textcolor{black}{Comparison with Different Path Planners.}}
	\label{tab:smallpath}
	\setlength{\tabcolsep}{5mm}
	\renewcommand\arraystretch{1.15}
	\resizebox{9cm}{!}{
		\color{black}
		\begin{tabular}{lcccc}
			\toprule
			\textbf{Method} & $E^{F}$ (Wh) & $StD$ & $T_{\text{tra}}$ (h) & $T_{\text{inf}}$ (s) \\
			\midrule
			ATOM~ & 16.42 & 0.35 & 8.56 &  3.57\\ 
			Pointerformer~ & 16.33 & 0.12 & 7.89 &  4.02\\ 
			AttRouting & 17.25 & 0.43 & 10.43 & \textbf{2.64} \\ 
			SLM (Ours) & \textbf{16.19} & \textbf{0.08} & \textbf{6.85} & 4.27 \\
			\bottomrule
		\end{tabular}
	}
\end{table}

%The proposed SLM attains the lowest flight energy consumption and the smallest standard deviation. This is because its geometry-aware path decoder effectively learns spatial regularities, while enhanced contextual queries and ensemble inference yield more stable policy gradients. In addition, SLM also achieves the fastest training speed, because it is fine-tuned on the pretrained BERT model and thus leverages existing linguistic knowledge, whereas the other models are trained from scratch. The embedded linguistic priors accelerate the convergence of the learning algorithm. By contrast, ATOM and Pointerformer can also reach near-optimal solutions but require substantially longer training time.

The proposed SLM attains the lowest flight energy consumption and the smallest standard deviation. This is because its geometry-aware path decoder effectively learns spatial regularities, while enhanced contextual queries and ensemble inference yield more stable policy gradients. \textcolor{black}{In addition, SLM also achieves the fastest training speed, because it is fine-tuned from a pretrained BERT backbone and thus benefits from the pretrained Transformer’s structural and optimization priors, whereas the other models are trained from scratch. These priors improve training stability and accelerate convergence}. 
\textcolor{black}{Among these path planners, SLM incurs the longest inference time, mainly because its geometric feature enhancement and ensemble inference introduce additional computational overhead during the inference stage.}
By contrast, ATOM and Pointerformer can also reach near-optimal solutions but require substantially longer training time.

\begin{figure}[htbp]
	\centering
	\includegraphics[width=9cm]{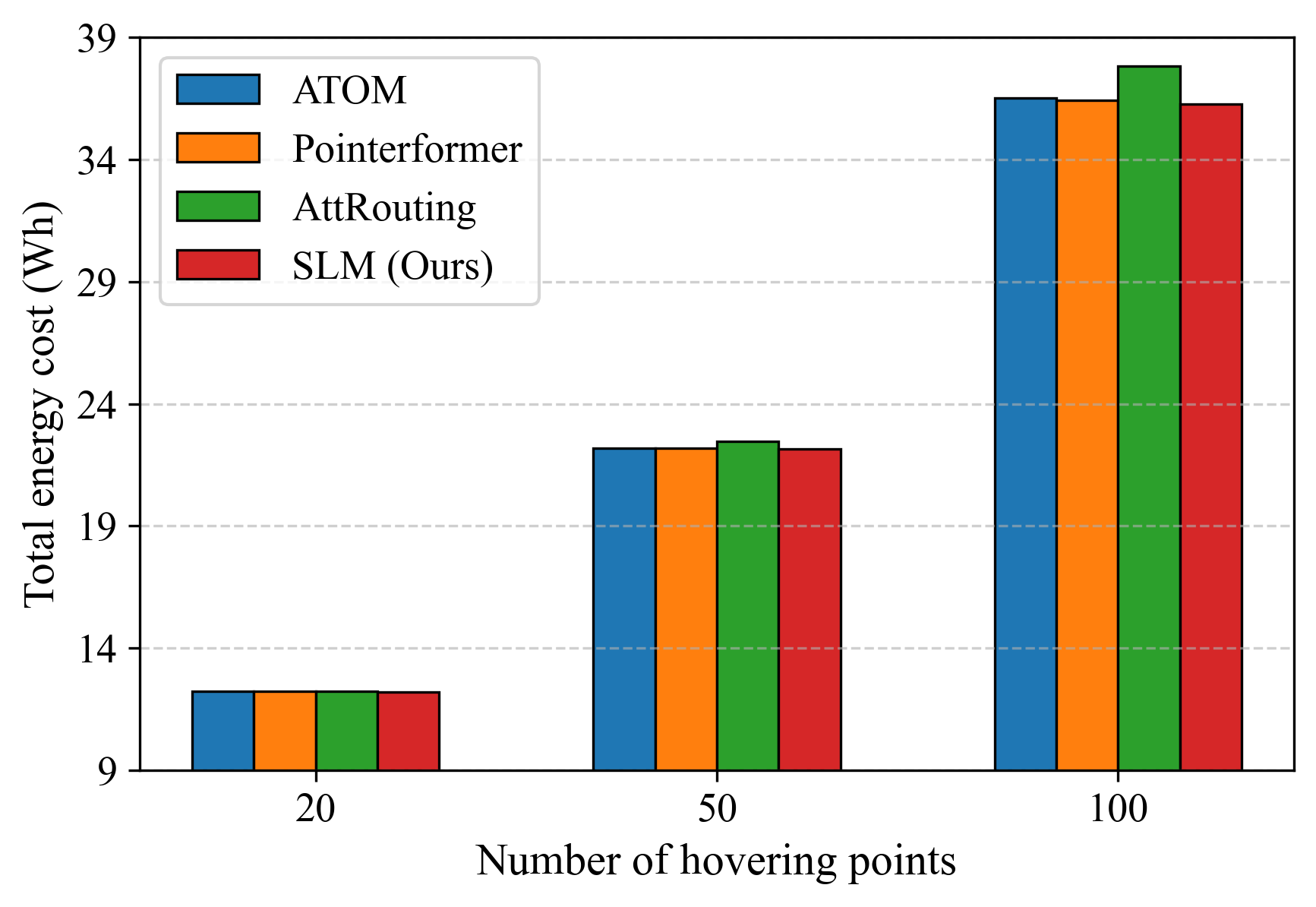}
	\caption{Energy Comparison of Lightweight Path Planners under Different Numbers of Hovering Points.}
	\label{fig:comparison1}
\end{figure}

We further compare the variations in total energy cost of different planners under varying numbers of hovering points. To ensure a fair comparison, the node distribution density is kept identical across all simulation scenarios. Fig. \ref{fig:comparison1} illustrates that the total energy cost of all planners increases with the number of hovering points. SLM consistently achieves the lowest energy consumption, while AttRouting yields the highest, and their gap widens as the problem scale grows, demonstrating SLM’s superior scalability.
\subsection{Experiment 3: Comparison of Different MLLMs}

To evaluate the reasoning capabilities of MLLMs for multi-UAV path planning tasks, we compare four representative models: GPT-4o, Gemini-2.5-Pro, Claude-Sonnet-4.0, and Mistral-Medium. Each model reasons over visualized coordinate maps to generate trajectories, followed by interactive refinement through human-in-the-loop prompting. The simulation area is set to 550 m × 550 m, containing 150 IoTD devices, and the number of UAV hovering points is defined as 30. The evaluation metrics include the total energy cost  $E_{\text{tot}}$, the flight energy consumption $E^{F} $,  and the inference time $T_{\text{inf}}$. The performance results of all MLLMs are listed in Table \ref{tab:mllm_comparison}.

\begin{table}[h]
	\centering
	\caption{Performance Comparison of Different MLLMs via Visual Reasoning.}
	\label{tab:mllm_comparison}
	\setlength{\tabcolsep}{5mm}
	\renewcommand\arraystretch{1.25}
	\resizebox{9cm}{!}{
		\begin{tabular}{lccc}
			\toprule
			\textbf{Model} & $E_{\text{tot}}$ (Wh) & $E^{F}$ (Wh) & $T_{\text{inf}}$ (s)\\
			\midrule
			Mistral-Medium     & 15.86 & 9.85 & \textbf{8.02} \\
			Gemini-2.5-Pro     & \textbf{13.66} & \textbf{7.64} & 55.25 \\
			Claude-Sonnet-4.0  & 14.57 & 8.58 & 9.63 \\
			GPT-4o             & 13.71 & 7.71 & 8.34 \\
			\bottomrule
	\end{tabular}}
\end{table}

Experimental results show that among all MLLMs, Gemini-2.5-Pro demonstrates stronger visual reasoning and achieves lower  energy consumption, albeit with a substantially longer inference time than its counterparts. GPT-4o delivers competitive performance with lower inference latency.

\begin{figure}[htbp]
	\centering
	\includegraphics[width=9cm]{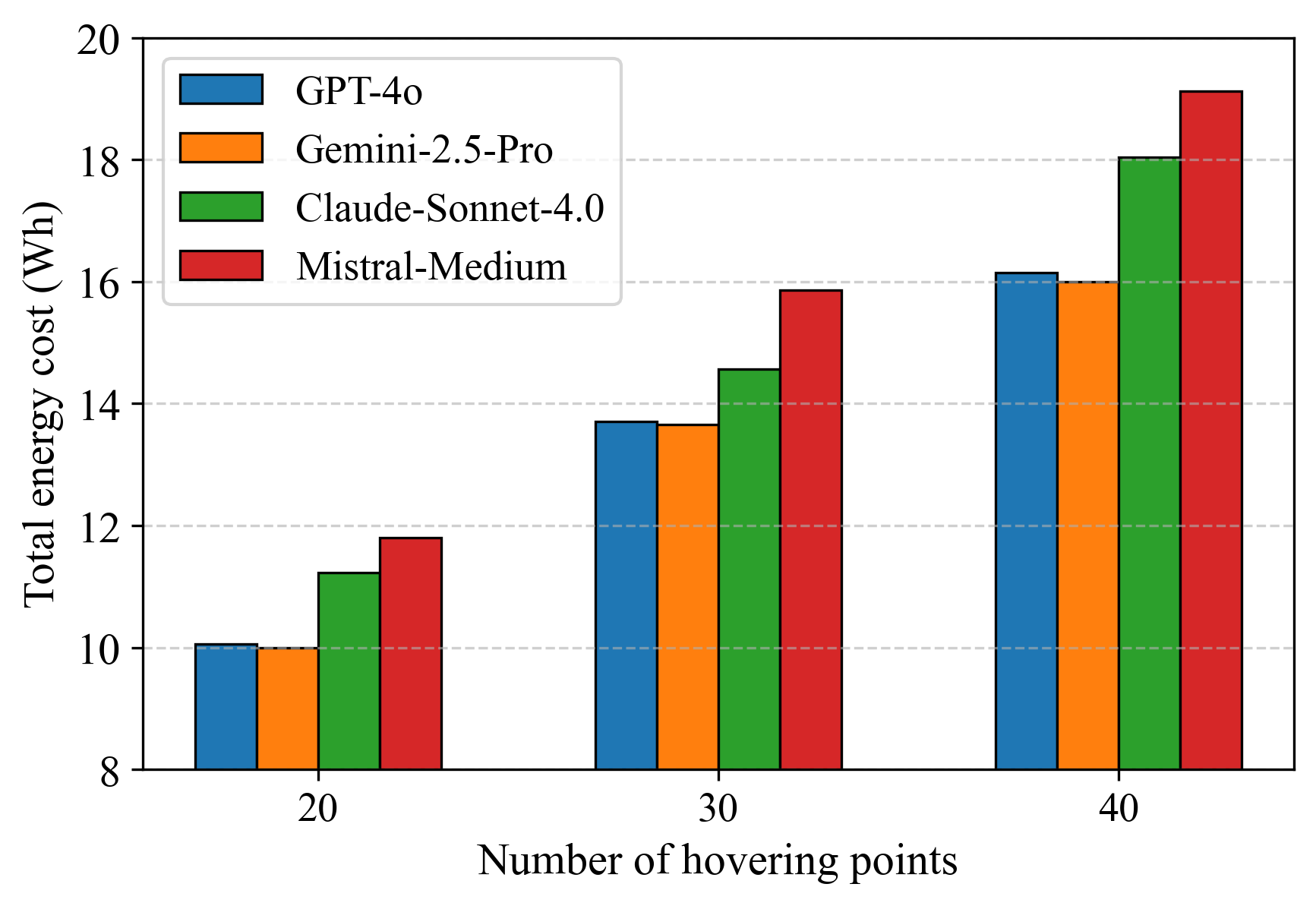}
	\caption{Energy Cost across MLLMs under Different Hovering Node Scales.}
	\label{fig:comparison2}
\end{figure}

Fig. \ref{fig:comparison2} compares the total energy consumption of four MLLMs under different numbers of hovering nodes. As the number of nodes increases, the energy consumption of all models rises due to the growth in trajectory length. Among them, Gemini-2.5-Pro consistently achieves the lowest energy consumption, reflecting its stronger visual reasoning and trajectory optimization capabilities, while GPT-4o exhibits more balanced overall performance.

\subsection{Experiment 4: Multi-Agent Collaboration Effectiveness}

This experiment aims to investigate the role of multi-agent collaboration within the Agentic AI framework for multi-UAV path planning, where GPT-4o is adopted as the underlying MLLM. Five configurations are evaluated: (A1) a single-agent system based on GPT-4o that autonomously refines trajectories through a ReAct mechanism \cite{10638533}, (A2) Initializer + Actor, (A3) Initializer + Actor + Critic, (A4) Initializer + Critic + Reflector, and (A5) the full framework incorporating four agents. The simulation area is set to 550 m × 550 m, containing 150 IoTD devices, and the number of UAV hovering points is defined as 30. All systems perform iterative optimization under the same experimental environment and are compared in terms of the total energy cost  $E_{\text{tot}}$, the constraint satisfaction rate $R_{\text{sat}}$,  and the inference time $T_{\text{inf}}$. We define the constraint satisfaction rate $R_{\text{sat}}$ as the fraction of planned paths that satisfy all feasibility constraints.
The performances of all Agent configurations are listed in Table \ref{tab:multiagent}.

\begin{table}[h]
	\centering
	\caption{Performance Comparison of Different Agent Configurations.}
	\label{tab:multiagent}
	\renewcommand\arraystretch{1.25}
	\resizebox{9cm}{!}{
		\begin{tabular}{lccc}
			\toprule
			\textbf{Configuration} & $E_{\text{tot}}$ (Wh) & $R_{\text{sat}}$ (\%) & $T_{\text{inf}}$ (s) \\
			\midrule
			A1: Single-agent + ReAct              & 13.68 & 20 & \textbf{74.52} \\
			A2: Initializer + Actor               & 12.95 & 27 & 138.26 \\
			A3: Initializer + Actor + Critic      & 12.72 & 72 & 105.25 \\
			A4: Initializer + Critic + Reflector  & 12.70 & 78 & 187.85 \\
			A5: Four-Agent                        & \textbf{12.43} & \textbf{83} & 116.47 \\
			\bottomrule
	\end{tabular}}
\end{table}

Experimental results show that the complete Agentic AI framework (A5) delivers the best overall performance: compared with other configurations, it achieves the highest constraint satisfaction rate and lower energy consumption. Specifically, the Critic introduces constraint-aware feedback via the MCP protocol to ensure solution feasibility; the Reflector provides reflective guidance that dynamically corrects suboptimal trajectories. The collaboration among these agents establishes an adaptive “generation-optimization–evaluation–reflection” loop, thereby enhancing the Agentic system’s intelligence, stability, and interpretability.

\begin{figure}[htbp]
	\centering
	\includegraphics[width=9cm]{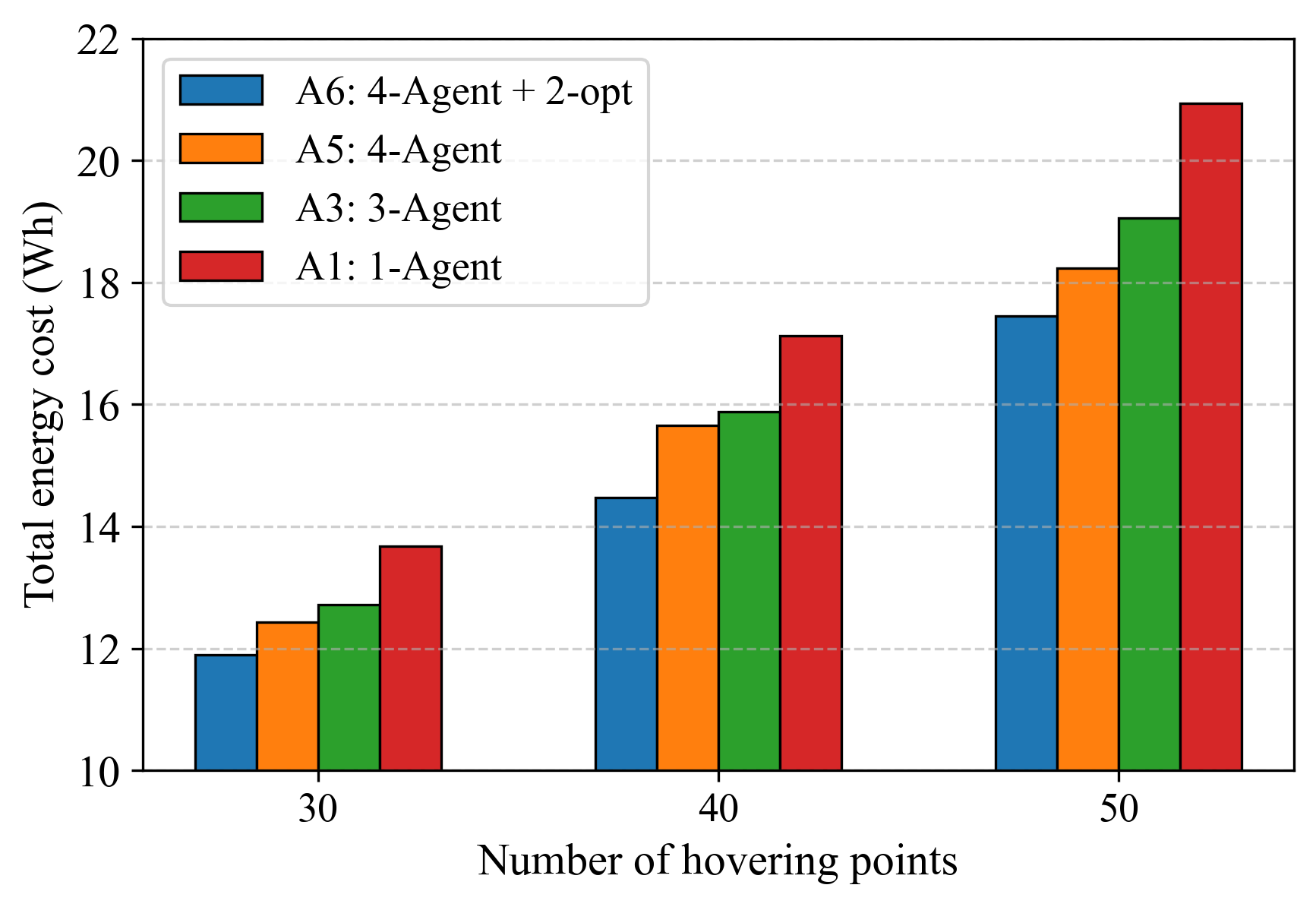}
	\caption{Energy Comparison of Agent Configurations under Different Numbers of Hovering Points.}
	\label{fig:comparison3}
\end{figure}

To overcome the limitations of pure visual reasoning and further enhance the performance of the Agentic AI system, we introduce a new agent configuration, A6, in which the Critic incorporates a 2-opt local search tool via the MCP Protocol.
Fig. \ref{fig:comparison3} compares the total energy consumption of different Agent configurations with 30, 40, and 50 hovering points. As the problem scale increases, all configurations exhibit a steady rise in energy consumption. 
The complete Agentic AI integrated with 2-opt local search consistently achieves the lowest energy cost across all scales, demonstrating the significant effectiveness of the MCP + tool mechanism in compensating for the limitations of visual reasoning and enhancing trajectory optimization performance.

\subsection{Experiment 5: Comparison among SLM, LLM, and Agentic AI}

A unified comparison is conducted across the following paradigms under varying problem scales in multi-UAV path planning scenarios: SLM, MLLM (GPT-4o), and Agentic AI. 
Each method is evaluated based on the trajectory length $L_{\text{traj}}$, the total energy cost  $E_{\text{tot}}$, the constraint satisfaction rate $R_{\text{sat}}$, and the inference time $T_{\text{inf}}$. 
Tables \ref{tab:finalcompare1} and \ref{tab:finalcompare2} report the performance comparison among SLM, LLM, and Agentic AI systems for 20 and 100 hovering points.
\begin{table}[h]
	\centering
	\caption{Comparison among SLM, LLM, and Agentic AI Systems with 20 Hovering Points.}
	\label{tab:finalcompare1}
	\renewcommand\arraystretch{1.25}
	\resizebox{9cm}{!}{
		\begin{tabular}{lcccc}
			\toprule
			\textbf{Method} & $L_{\text{traj}}$ (m) & $E_{\text{tot}}$ (Wh) & $R_{\text{sat}}$ (\%) & $T_{\text{inf}}$ (s) \\
			\midrule
			SLM                 & 5428.75 & 15.56 & \textbf{95} & \textbf{3.48} \\
			MLLM (GPT-4o)       &5685.52 & 16.09 & 33 & 8.62 \\
			Agentic AI          & 5130.44 & 14.94 & 85 & 87.85 \\
			Agentic AI + 2-opt   & \textbf{4540.12} & \textbf{13.71} & 85 & 112.47 \\
			\bottomrule
	\end{tabular}}
\end{table}
\begin{table}[h]
	\centering
	\caption{Comparison among SLM, LLM, and Agentic AI Systems with 100 Hovering Points.}
	\label{tab:finalcompare2}
	\renewcommand\arraystretch{1.25}
	\resizebox{9cm}{!}{
		\begin{tabular}{lcccc}
			\toprule
			\textbf{Method} & $L_{\text{traj}}$ (m) & $E_{\text{tot}}$ (Wh) & $R_{\text{sat}}$ (\%) & $T_{\text{inf}}$ (s)\\
			\midrule
			SLM                 & \textbf{8938.45} & \textcolor{black}{\textbf{39.72}} & \textbf{92} & \textbf{4.35} \\
			MLLM (GPT-4o)       & 10865.26 & 42.72 & 12 & 9.56 \\
			Agentic AI          & 10610.21 & 42.18 & 68 & 384.65 \\
			Agentic AI + 2-opt   & 9468.36 & 39.83 & 68 & 443.27 \\
			\bottomrule
	\end{tabular}}
\end{table}

The results indicate that, in small-scale system optimization, both SLM and Agentic AI achieve low energy consumption; in particular, the Agentic AI system augmented with 2-opt local search attains performance that is nearly at the theoretical optimum. Because constraint information is incorporated during training, SLM sustains the highest constraint satisfaction rate, and its inference latency is extremely low—substantially lower than that of MLLMs and Agentic AI systems. In large-scale system optimization, the performance of vision-reasoning–based MLLMs and Agentic AI degrades markedly, whereas SLM delivers the best overall results, achieving the highest constraint satisfaction rate and the lowest inference latency. Although integrating 2-opt into the Agentic AI pipeline can compensate for the limitations of visual reasoning and yield competitive performance, it comes at the cost of a significant increase in inference latency.

\subsection{Summary and Lessons Learned}

\begin{figure}[htbp]
	\centering
	\includegraphics[width=8cm]{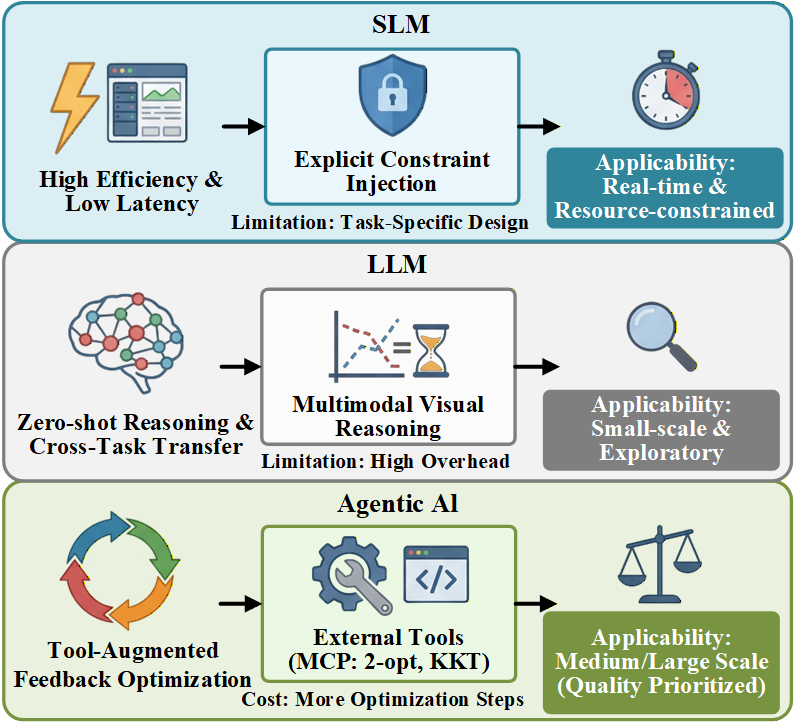}
	\caption{\textcolor{black}{Comparison of SLM, LLM, and Agentic AI Paradigms.}}
	\label{fig:bijiao}
\end{figure}

\textcolor{black}{Under comprehensive and rigorous evaluation, we summarize the characteristics and applicability of SLM, LLM, and Agentic AI for energy-efficient UAV-enabled WPT systems as follows, with the overall comparison illustrated in Fig. \ref{fig:bijiao}.}

\subsubsection{SLM}  Given comparable hardware and task scales, SLM attains the highest computational efficiency and the lowest inference latency, and achieves the best energy performance even in large-scale system optimization. Because constraint knowledge is explicitly injected during training, SLM typically yields the highest constraint satisfaction rate and more stable policy gradients. Its limitation lies in task specificity and limited cross-task generalization, necessitating retraining for different optimization objectives or constraint sets. %In addition, the linguistic knowledge embedded in SLM accelerates training. 

\textit{Applicability}: real-time, resource-constrained deployments across varying problem scales, or scenarios with stable objectives/constraints that allow offline retraining.

\subsubsection{LLM} LLMs exhibit strong zero-shot reasoning and generalization in small-scale optimization problems, enabling rapid validation and transfer across tasks. However, for large-scale optimization problems (e.g., path planning), they incur substantial computational overhead and higher latency, with degraded accuracy—reflecting current bottlenecks in visual reasoning and long-horizon global optimization. 

\textit{Applicability}: small-scale, exploratory tasks requiring quick adoption and cross-domain generalization; less suitable for stringent real-time constraints and severely limited computational resources.

\subsubsection{Agentic AI} Agentic AI offers a well-balanced overall profile by enabling self-correction and adaptive planning via the “generation-optimization–evaluation–reflection” loop. In large-scale optimization problems, integrating external tools via MCP (e.g., 2-opt and KKT) can markedly compensate for MLLM’s limitations in visual reasoning, driving the energy/trajectory toward near-optimality—at the cost of increased inference latency and more optimization steps. 

\textit{Applicability}: medium-to-large problem sizes where feasibility, robustness, and interpretability are prioritized, and where some latency can be traded for higher solution quality.

\section{Conclusions}
\label{sec:conclusion}

This study investigates the advantages and applicability of SLMs, LLMs, and Agentic AI for UAV-enabled WPT systems. First, we propose a lightweight BERT-based framework with enhanced UAV embeddings, contextual features, a geometry-aware path decoder, and ensemble inference for efficient path generation and time allocation under limited computation and energy resources. Second, to address dynamic environments and multiple constraints, we develop an Agentic AI framework to exploit the reasoning and interactive capabilities of LLMs. Through a closed-loop process of path generation, optimization, evaluation, and reflection, the system enables more adaptive and intelligent decision-making. Experimental results show that SLMs achieve low latency and high energy efficiency in resource-constrained scenarios, whereas LLMs and Agentic AI provide stronger generalization and decision-making capability in cross-task planning.

\textcolor{black}{
Future work can further enhance UAV-enabled WPT systems in several directions. First, model lightweighting and compression can reduce deployment overhead on edge or airborne platforms. Techniques such as model distillation, lightweight LLMs (e.g., LLaMA2-7B), and edge-cloud collaboration can further lower inference cost. Second, expanding tool-chain integration in Agentic AI frameworks may improve optimization performance. Third, stronger cross-scenario generalization and extension to mobile IoTD scenarios will improve adaptation to more dynamic UAV-enabled WPT environments. Finally, practical engineering issues, including communication latency, on-device inference power, privacy, and security, should be addressed to strengthen real-world applicability.}

\bibliographystyle{IEEEtran}
\bibliography{bare_jrnl}

\vspace{-1.2cm}
\begin{IEEEbiography}[{\includegraphics[width=1in,height=1.25in,clip,keepaspectratio]{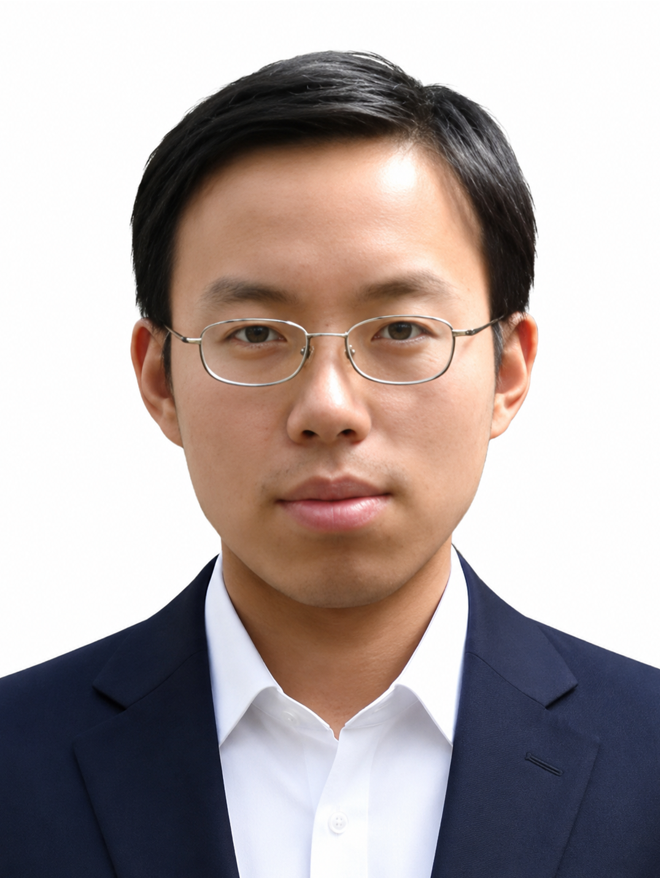}}]{Feibo Jiang} received the B.S. and M.S. degrees from the School of Physics and Electronics, Hunan Normal University, China, in 2004 and 2007, respectively, and the Ph.D. degree from the School of Geosciences and Info-Physics, Central South University, China, in 2014. He is currently an Associate Professor with the Hunan Provincial Key Laboratory of Intelligent Computing and Language Information Processing, Hunan Normal University, China. His research interests include large model–assisted wireless communications, Agentic AI–enabled wireless communications, semantic communications, and UAV networks. He serves as a (Lead) Guest Editor for IEEE Journal on Selected Areas in Communications, IEEE Wireless Communications, IEEE Network, and IEEE Communications Magazine.
\end{IEEEbiography}
\vspace{-1.2cm}
\begin{IEEEbiography}[{\includegraphics[width=1in,height=1.25in,clip,keepaspectratio]{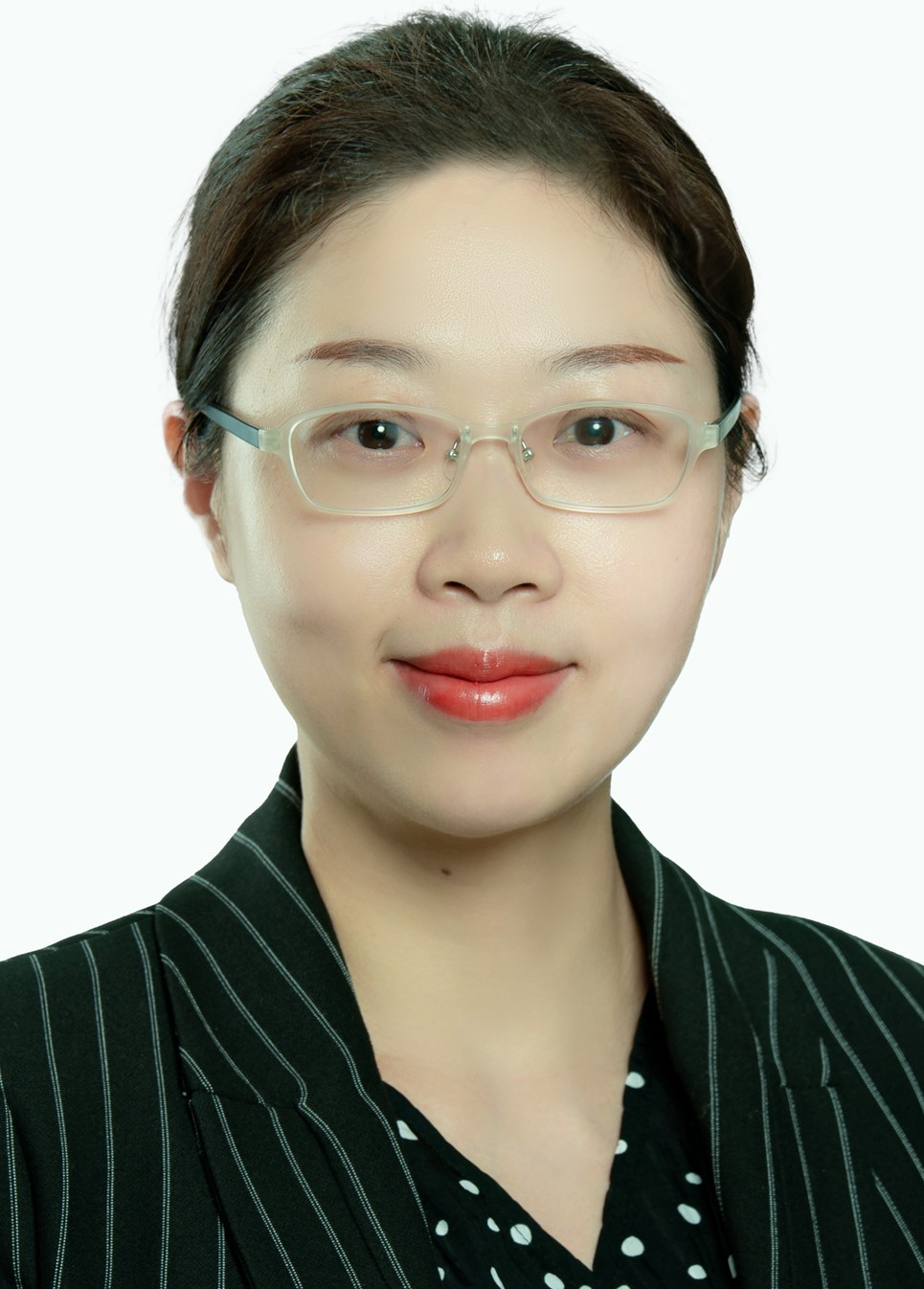}}]{Li Dong} received the B.S. and M.S. degrees in School of Physics and Electronics from Hunan Normal University, China, in 2004 and 2007, respectively. She received her Ph.D. degree in School of Geosciences and Info-physics from the Central South University, China, in 2018. She is currently a professor at Hunan University of Technology and Business, China. Her research interests include machine learning, Internet of Things, and mobile edge computing.
\end{IEEEbiography}
\vspace{-1.2cm}
\begin{IEEEbiography}[{\includegraphics[width=1in,height=1.25in,clip,keepaspectratio]{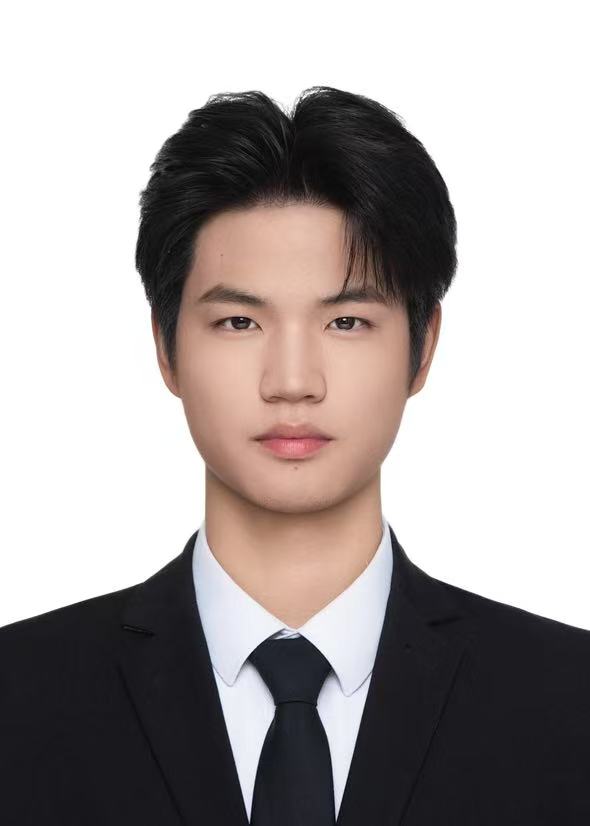}}]{Lei Mao} is currently pursuing the M.S. degree with the School of Computer Science and Technology, Hunan Normal University. His main research interests include semantic communication and Agentic AI-enabled wireless communications.
\end{IEEEbiography}
\vspace{-1.2cm}
\begin{IEEEbiography}[{\includegraphics[width=1in,height=1.25in,clip,keepaspectratio]{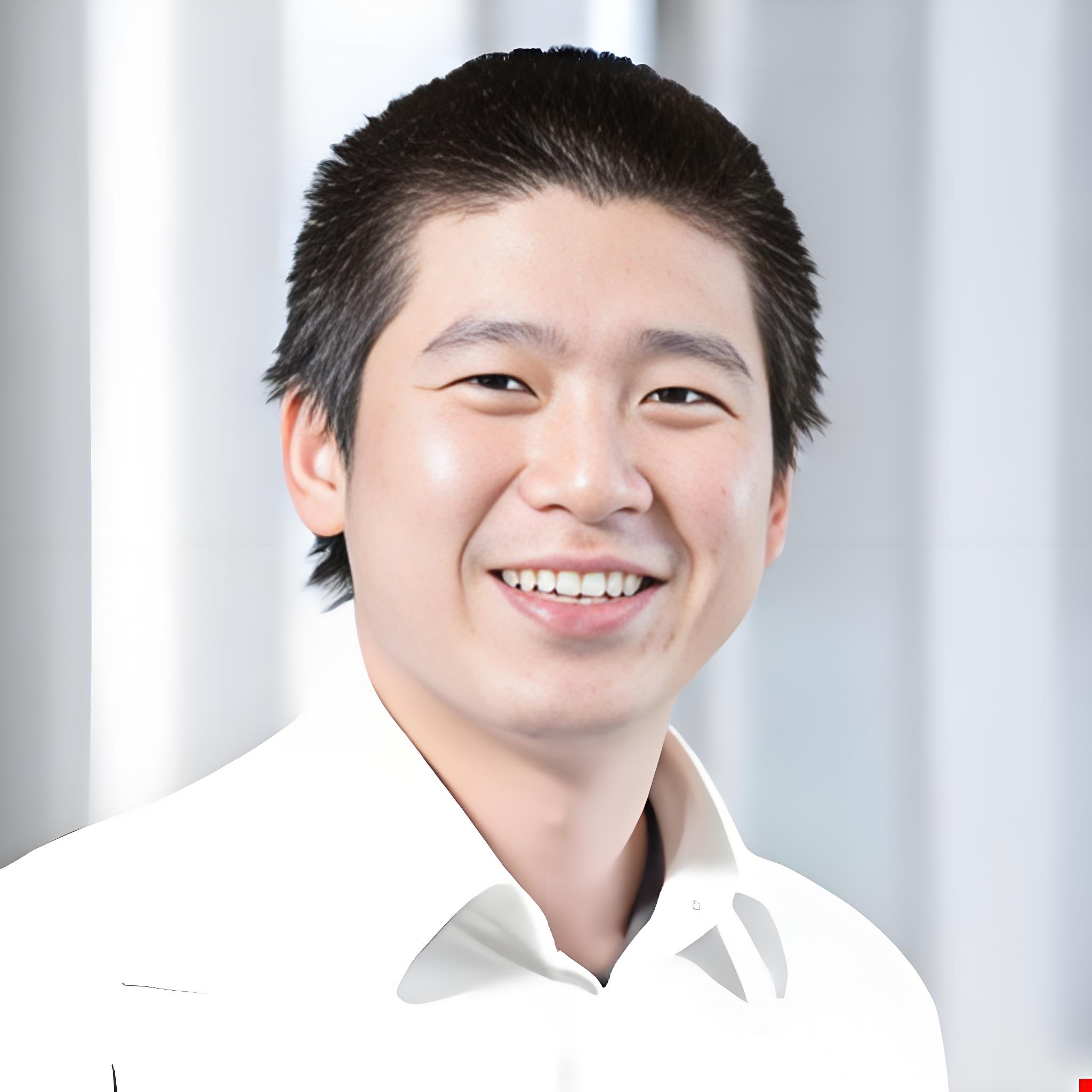}}]{Kezhi Wang} received the Ph.D. degree from the University of Warwick, U.K., funded by the Chancellor’s International Scholarship. He is currently a Professor with the Department of Computer Science, Brunel University London, U.K. He is also a Royal Society Industry Fellow and has been recognised as a Highly Cited Researcher by Clarivate Web of Science and a Top 2\% Scientist of the World. He has published over 150 papers in IEEE journals and has received several prestigious awards, including the IEEE Communications Society Leonard G. Abraham Prize, Heinrich Hertz Award, and Fred W. Ellersick Prize. His research interests include semantic communications, machine learning for communication systems, mobile edge computing, and wireless networks. He serves as an Associate Editor for IEEE Transactions on Vehicular Technology, IEEE Wireless Communications Letters, and IEEE Communications Letters, and has been actively involved in numerous IEEE conferences and editorial roles.  
\end{IEEEbiography}
\vspace{-1.2cm}
\begin{IEEEbiography}[{\includegraphics[width=1in,height=1.25in,clip,keepaspectratio]{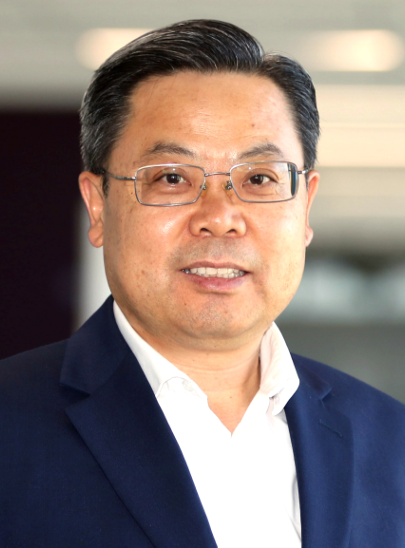}}]{Xianbin Wang} received his Ph.D. degree in electrical and computer engineering from the National University of Singapore in 2001. He has been with Western University, Canada, since 2008, where he is currently a Distinguished University Professor and Tier-1 Canada Research Chair in Trusted Communications and Computing. Prior to joining Western University, he was with the Communications Research Centre Canada as a Research Scientist and later a Senior Research Scientist from 2002 to 2007. From 2001 to 2002, he was a System Designer at STMicroelectronics. His current research interests include 5G/6G technologies, Internet of Things, machine learning, communications security, digital twin, and intelligent communications. He has over 800 highly cited journals and conference papers, in addition to over 30 granted and pending patents and several standard contributions.
	
	Dr. Wang is a Fellow of the Canadian Academy of Engineering and a Fellow of the Engineering Institute of Canada. He has received many prestigious awards and recognitions, including the IEEE Canada R. A. Fessenden Award, Canada Research Chair, Engineering Research Excellence Award at Western University, Canadian Federal Government Public Service Award, Ontario Early Researcher Award, and twelve Best Paper Awards. He is currently a member of the Senate, Senate Committee on Academic Policy and Senate Committee on University Planning at Western. He also serves on NSERC Discovery Grant Review Panel for Computer Science. He has been involved in many flagship conferences, including IEEE GLOBECOM, ICC, VTC, PIMRC, WCNC, CCECE, and ICNC, in different roles, such as General Chair, TPC Chair, Symposium Chair, Tutorial Instructor, Track Chair, Session Chair, and Keynote Speaker. He was nominated as an IEEE Distinguished Lecturer multiple times by different societies including BTS, ComSoc and VTS. He serves/has served as the Editor-in-Chief, Associate Editor-in-Chief, Area Editor, and editor/associate editor for over ten journals. He was the Chair of the IEEE ComSoc Signal Processing and Computing for Communications (SPCC) Technical Committee and the Central Area Chair of IEEE Canada.
	
\end{IEEEbiography}
\vspace{-1.2cm}
\begin{IEEEbiography}[{\includegraphics[width=1in,height=1.25in,clip,keepaspectratio]{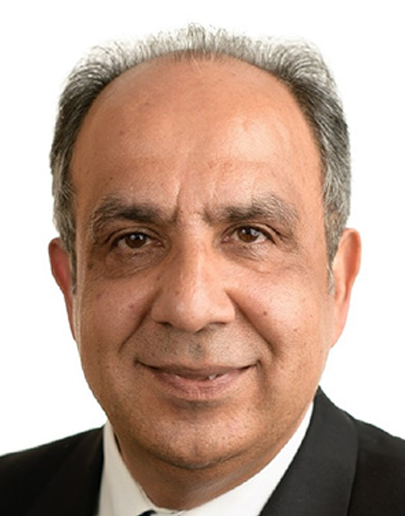}}]{Abbas Jamalipour} (S’86–M’91–SM’00–F’07) received the Ph.D. degree in Electrical Engineering from Nagoya University, Nagoya, Japan in 1996. He is the Professor of Ubiquitous Mobile Networking at the University of Sydney. He has authored nine technical books, eleven book chapters, over 650 technical papers, and five patents, all in wireless communications and networking. Prof. Jamalipour is the recipient of several prestigious awards such as the 2026 IEEE James Evans Avant Garde Award, Engineers Australia 2025 Neville Thiele Eminence Award, IEEE ComSoc Harold Sobol Award, the IEEE ComSoc Best Tutorial Paper Award, as well as over fifteen Best Paper Awards. He is currently serving as the IEEE ComSoc Vice President-MGA and was the Editor-in-Chief IEEE WIRELESS COMMUNICATIONS, Vice President-Conferences, and a member of Board of Governors of the IEEE Communications Society. Within IEEE Vehicular Technology Society, he has held positions of President, Executive Vice-President, elected member of the Board of Governors, and the Editor-in-Chief of IEEE TRANSACTIONS ON VEHICULAR TECHNOLOGY and VTS Mobile World. He is a Fellow of the Institute of Electrical, Information, and Communication Engineers (IEICE), the Institution of Engineers Australia (IEAust), and the International Artificial Intelligence Industry Alliance (AIIA), and a Visiting Fellow of the UK Royal Academy of Engineering.
	
\end{IEEEbiography}
%\bibliography{bare_jrnl}

\newpage
\end{document}